%% file: gj1214v2.tex
\documentclass[twocolumn,twocolumnappendix]{aastex63}
\usepackage{amsmath}

\newcommand{\teff}[1]{$T_{\text{eff}}$#1}
\newcommand{\prot}[1]{$P_{\rm rot}$#1}

\newcommand{\logg}[1]{$\log{g}$#1}

\newcommand{\msini}[1]{$m_p\sin{i}$#1}
\newcommand{\mps}[1]{m s$^{-1}$#1}
\newcommand{\teq}[1]{$T_{\text{eq}}$#1}
\newcommand{\Rearth}[1]{$R_{\oplus}$#1}
\newcommand{\Rsun}[1]{$R_{\odot}$#1}
\newcommand{\Mearth}[1]{$M_{\oplus}$#1}
\newcommand{\Msun}[1]{$M_{\odot}$#1}
\newcommand{\name}[1]{GJ 1214#1}

\newcommand{\Krv}[1]{$14.36\pm 0.53$#1}
\newcommand{\ecc}[1]{$0.063$#1}
\newcommand{\eccnn}[1]{$0.100$#1}
\newcommand{\mplanet}[1]{$8.17\pm 0.43$#1}
\newcommand{\rplanett}[1]{$2.733^{+0.050}_{-0.052}$#1}
\newcommand{\rplanetf}[1]{$2.749^{+0.050}_{-0.054}$#1}
\newcommand{\rplanet}[1]{$2.742^{+0.050}_{-0.053}$#1}
\newcommand{\rhoplanet}[1]{$2.20^{+0.17}_{-0.16}$#1}
\newcommand{\porb}[1]{$1.58040433\pm 0.00000013$#1}
\newcommand{\sma}[1]{$0.01490\pm 0.00026$#1}
\newcommand{\insol}[1]{$21.0^{+2.7}_{-2.5}$#1}
\newcommand{\teqval}[1]{$596\pm 19$#1}
\newcommand{\Xenv}[1]{$5.24^{+0.30}_{-0.29}$#1}
\newcommand{\Nrv}[1]{$15$#1}
\newcommand{\Ntransit}[1]{$12$#1}
\newcommand{\protval}[1]{$124.7^{+5.0}_{-4.8}$#1}
\newcommand{\ftransitclean}[1]{$86^{+6}_{-20}$\%#1}  
\newcommand{\ftransitminuskep}[1]{$90^{+5}_{-21}$\%#1}  
\newcommand{\frvclean}[1]{$75^{+9}_{-12}$\%#1}  
\newcommand{\frvminuskep}[1]{$85^{+8}_{-15}$\%#1}  
\newcommand{\fonep}[1]{$44^{+9}_{-5}$}  
\newcommand{\fallrv}[1]{$12$\%}  

\defcitealias{anglada13b}{An13}
\defcitealias{anglada14}{An14}
\defcitealias{anglada16}{An16}
\defcitealias{astudillodefru17c}{As17b}
\defcitealias{astudillodefru17a}{As17a}
\defcitealias{bonfils07}{Bo07}
\defcitealias{bonfils18a}{Bo18a}
\defcitealias{bonfils18b}{Bo18b}
\defcitealias{cloutier19c}{Cl19}
\defcitealias{cloutier20b}{Cl20}
\defcitealias{diaz19}{D\'{i}19}
\defcitealias{dreizler20}{Dr20}
\defcitealias{feng20a}{Fe20a}
\defcitealias{feng20b}{Fe20b}
\defcitealias{kemmer20}{Ke20}
\defcitealias{lillobox20}{Li20}
\defcitealias{luque18}{Lu18}
\defcitealias{luque19a}{Lu19}
\defcitealias{ment19}{Me19}
\defcitealias{nowak20}{No20}
\defcitealias{pinamonti18}{Pi18}
\defcitealias{shporer20}{Sh20}
\defcitealias{soto21}{So21}
\defcitealias{stock20}{St20}
\defcitealias{suarez17}{Su17}
\defcitealias{suarez20}{Su20}
\defcitealias{trifonov18}{Tr18}
\defcitealias{trifonov21}{Tr21}
\defcitealias{tuomi14}{Tu14}
\defcitealias{vaneylen21}{Va21}
\defcitealias{winters21}{Wi21}

\shortauthors{Cloutier et al.}
\shorttitle{}

\begin{document}
\title{A More Precise Mass for GJ 1214 b and the Frequency of Multi-Planet Systems Around Mid-M Dwarfs}

\input{allauthors}

\correspondingauthor{Ryan Cloutier}
\email{ryan.cloutier@cfa.harvard.edu}

\begin{abstract}
  We present an intensive effort to refine the mass and orbit
  of the enveloped terrestrial planet \name{} b using 165 radial velocity (RV)
  measurements taken with the HARPS spectrograph over a period of ten
  years. We conduct a joint analysis of the RVs with archival Spitzer/IRAC
  transits and measure a planetary mass and radius of \mplanet{} \Mearth{}
  and \rplanet{} \Rearth{.} Assuming GJ 1214 b is an Earth-like core surrounded
  by a H/He envelope, we measure an envelope mass fraction of
  $X_{\rm env}=$ \Xenv{\%}.
  \name{} b remains a prime target for secondary eclipse observations
  of an enveloped terrestrial, the scheduling of which benefits from
  our tight constraint on the orbital eccentricity of $<$ \ecc{} at 95\%
  confidence,
  which narrows the secondary eclipse window to 2.8 hours. By combining \name{}
  with other mid-M dwarf transiting systems with intensive RV follow-up, we
  calculate the frequency of mid-M dwarf planetary systems with multiple small
  planets and find
  that \ftransitminuskep{} of mid-M dwarfs with a known planet with mass
  $\in [1,10]$ \Mearth{} and orbital period $\in [0.5,50]$ days, will host at
  least one additional planet. We rule out additional planets around \name{}
  down to
  3 \Mearth{} within 10 days such that \name{} is a single-planet system within
  these limits, a result that has a \fonep{}\% probability given the
  prevalence of multi-planet systems around mid-M dwarfs. We also investigate
  mid-M dwarf RV systems and show that the probability that all reported RV
  planet candidates are real planets is $<12$\% at 99\% confidence, although
  this statistical argument is unable to identify the probable false positives.
\end{abstract}


\section{Introduction}
\name{} remains a benchmark system for M dwarf planetary systems with small
planets in the super-Earth to sub-Neptune size regime \citep{charbonneau09}.
The accessibility of the \name{} system for detailed characterization has
warranted measurements of spectroscopic stellar parameters \citep{anglada13a}, 
photometric variability studies \citep{berta11,narita13,nascimbeni15},
and X-ray activity monitoring
\citep{lalitha14}. The planet \name{} b itself has been subject to a multitude 
of searches for transit timing variations (TTV)
\citep{carter11,harpsoe13,fraine13,gillon14} and atmospheric characterization
efforts
\citep{bean11,croll11,crossfield11,demooij12,fraine13,kreidberg14a}.
The TTV searches have sought to 1) scrutinize the multiplicity of the
inner planetary system, and 2) to improve constraints on the mass of \name{} b
if a second transiting planet was uncovered. These searches yielded null
results and consequently were unable to provide an independent mass measurement
of \name{} b. Years of follow-up campaigns have generated the consensus that
\name{} is an inactive mid-M dwarf that hosts a single known planet,
and given the radial velocity (RV) mass and
radius of \name{} b ($m_p=6.26\pm 0.91$ \Mearth{;} \citealt{anglada13a},
$2.85\pm 0.2$ \Rearth{;} \citealt{harpsoe13}), the planet must be enveloped in
H/He gas, although high-altitude clouds hinder the detection of chemical
species in the deep atmosphere \citep{kreidberg14a}.

The null results from the aforementioned TTV searches and the absence of
additional RV measurements analyzed since the planet's discovery implies
that the mass of \name{} b has not been refined since its discovery.
\cite{anglada13a} did however conduct a stellar plus RV analysis using the
same RV data from the discovery paper \citep{charbonneau09}. \cite{anglada13a}
performed an independent RV extraction using the \texttt{TERRA}
template-matching algorithm \citep{anglada12} and recovered a consistent RV
semiamplitude with an equivalent measurement uncertainty of $\pm 1.6$ \mps{.}
 
In this study, we analyze the ten years of intensive RV follow-up of \name{}
with the aim to improve our understanding of the mass and orbital solution of
\name{} b and to search for additional planetary signals in the RV data. With
these RV data, \name{} joins the set of mid-M dwarfs with sufficient RV
observations to be sensitive to the detection of additional sub-Neptune-mass
planets within the system. By analyzing the RV sensitivity of \name{} along
with other mid-M dwarf planetary systems with comparable RV follow-up datasets,
we present the frequency of mid-M dwarf planetary systems with multiple small
planets with masses between 1-10 \Mearth{} and with orbital periods between
0.5-50 days.

In Section~\ref{sect:star} we report the adopted stellar parameters.
In Section~\ref{sect:obs} we present the RV and transit observations used in
this study. In Section~\ref{sect:analysis} we jointly model the RV and transit
data to measure the mass and orbit of \name{} b. In Section~\ref{sect:sens} we
derive the RV sensitivity and place constraints on the presence of additional
planets orbiting \name{.} In Section~\ref{sect:freq} we compute the frequency
the multi-planet systems orbiting mid-M dwarfs. We conclude with a summary of
our results in Section~\ref{sect:summary}.

\section{Stellar Parameters} \label{sect:star}
GJ 1214 is a metal-rich M4 dwarf ([Fe/H]$ = 0.29\pm 0.12$ dex;
\citealt{newton14}) located at a distance of
$14.631 \pm 0.012$ pc \citep{lindegren21,bailer21}. We use the star's absolute
$K_s$-band magnitude ($M_{K}=7.956 \pm 0.020$) and metallicity to infer its
stellar mass ($M_s =0.178\pm 0.010$ \Msun{)} and radius
($R_s =0.215\pm 0.008$ \Rsun{)} using the empirical M dwarf mass-luminosity and
radius-luminosity relations from \cite{benedict16} and \cite{mann15},
respectively. We adopt the stellar effective temperature from
\cite{anglada13a}, which derived \teff{} via SED fitting to the absolute
$BVRIJHK_{s}$W1W2W3W4 magnitudes. The resulting value of \teff{} $=3250\pm 100$
K is in good agreement with the values derived from $K$-band water absorption
\citep{rojasayala12} and from empirical temperature-color relations
\citep{mann15}. The stellar luminosity, density, and surface gravity are 
derived self-consistently from the aforementioned parameters.
The stellar parameters adopted throughout this work are
summarized in Table~\ref{tab:star}.

\input{gj1214table}

\section{Observations} \label{sect:obs}
\subsection{HARPS Radial Velocities} \label{sect:harps}
\name{} b was observed with the High Accuracy Radial velocity Planet Searcher
\citep[HARPS;][]{mayor03} \'echelle spectrograph mounted at the 3.6m ESO
telescope at La Silla Observatory, Chile. A total of 168 spectra were
taken between UT 2009 June 11 and UT 2019 September 2, although we exclude
three spectra with exceptionally low S/N (i.e. total S/N $<5$).
Included in the remaining set of 165 spectra are the 21 RV observations
originally published in the discovery paper of \name{} b \citep{charbonneau09}
and subsequently reanalyzed by \cite{anglada13a}.
The observations were obtained through five time allocations: ESO
programs 183.C-0972, 283.C-5022 (P.I. Udry), and 1102.C-0339, 183.C-0437,
198.C-0838 (P.I. Bonfils). The exposure time was set to 2400 seconds in each
program.
Due to the low flux of the target spectrum, the simultaneous cailbration
fiber was set to on-sky rather than to the standard thorium argon lamp to
avoid any potential contamination of the lamp into the science fiber at the
bluest \'echelle orders. A majority of the spectra were
also read in the HARPS slow mode in order to minimize the read-out noise.
In the first two observing programs (22/165 observations),
we obtained a median S/N at the center of the order 66 ($\sim 650$ nm)
of 18.8, which corresponds to a median RV uncertainty of 2.35 \mps{.}
In the remaining ESO programs 1102.C-0339 (10/165 observations), 183.C-0437
(68/165 observations), and 198.C-0838 (65/165 observations), median
S/N values at 650 nm of 13.9, 16.6, and 16.2, respectively, were obtained.
The corresponding median RV uncertainties were 3.81, 3.35, and 3.25
\mps{.}

To achieve the RV uncertainties quoted above, we extracted each spectrum's
RV using the \texttt{TERRA}
pipeline \citep{anglada12}. \texttt{TERRA} employs a template-matching
scheme which has been shown to produce an improved RV measurement precision
on M dwarfs compared to the cross-correlation function (CCF) technique
\citep[e.g.][]{anglada12,astudillodefru15}, which is employed
by the HARPS Data Reduction Software \citep[DRS;][]{lovis07}.
\texttt{TERRA} constructs an empirical, high S/N template spectrum by coadding
the individual spectra after shifting to the solar system
barycentric reference frame using the
barycentric corrections calculated by the HARPS DRS. Spectral regions with
telluric absorption exceeding 1\% are omitted from the RV extraction process. 
Each spectrum's RV is extracted via a linear least squares fit to the master
template in velocity space. In our analysis, we ignore the bluest orders at low
S/N and only consider spectral orders redward of
order number 139 (440-687 nm).

In June 2015 (BJD = 2457176.5), the HARPS fiber link was upgraded to improve the
instrument's throughput, stability, and the uniformity of the emergent
illumination pattern \citep{locurto15}. These improvements enhanced the RV
stability of the instrument but produced a velocity offset that we account for
in the RV extraction by constructing two template spectra for the subsets of
data corresponding to pre and post-fiber upgrade (containing 90 and 75 out of
165 measurements, respectively). The resulting raw RV time series is provided in
Table~\ref{tab:rv} and does not include corrections for the secular
acceleration of GJ 1214\footnote{The parallax and proper motion of GJ 1214 from Gaia EDR3 produce an instantaneous secular acceleration at J2015.5 of $\text{d}v_r / \text{d}t = 0.3$ m/s/yr \citep[Eq. 2;][]{zechmeister09b}.}, nor for any
Rossiter-induced contamination of in-transit measurements. We find that six out
of 165 of our RV measurements occur in-transit and may be marginally affected by
the Rossiter-McLaughlin (RM) effect. Due to the slow rotation of GJ 1214, we can
safely ignore this effect in our RV modeling as the expected RM signal
amplitude is neglible compared to our typical RV uncertainties
\citep[$K_{\rm RM} \sim 1.1$ \mps{;}][]{gaudi07} and in practice would
appear smaller given that the exposure time is comparable to the transit
duration.

\input{rvs}

\begin{figure*}
  \centering
  \includegraphics[width=.9\hsize]{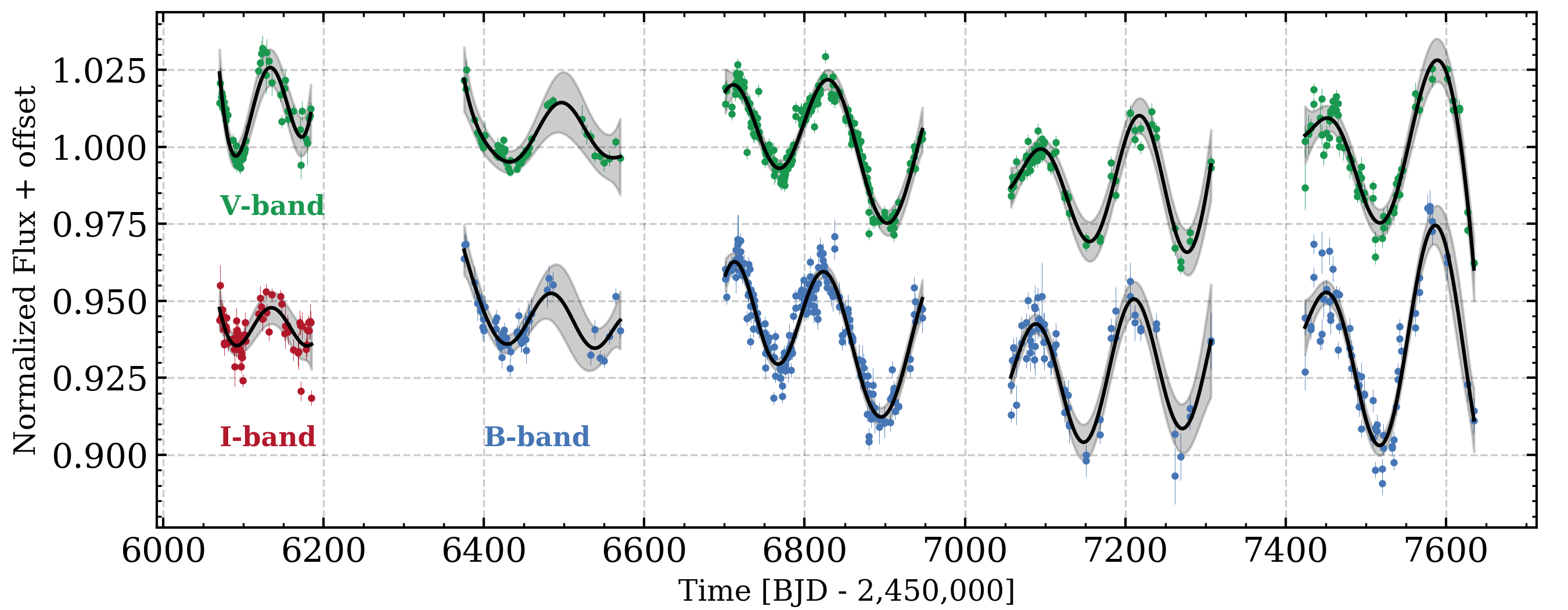}
  \caption{Long-term multi-color photometric monitoring of \name{} with STELLA
    photometry. The blue, green, and red markers depict the $BVI$ light curves,
    respectively. The black curves depict the mean function of the
    quasi-periodic Gaussian process while the surrounding shaded region depicts
    its $3\sigma$ uncertainty.}
  \label{fig:prot}
\end{figure*}

\subsection{Spitzer/IRAC Transit Photometry} \label{sect:spitzer}
\cite{fraine13} and \cite{gillon14} presented data from a continuous
photometric monitoring program of \name{} with the
\text{Spitzer Space Telescope}. The program's strategy and motivation were to
monitor \name{} for 20.9 days to characterize the infrared transit and secondary
eclipse of \name{} b, and to search for additional transiting planets down to
Martian size out to the inner edge of the star's habitable zone. These
observations did not yield a significant detection of the secondary eclipse
depth, nor did they yield any additional planet detections. However, these
data provide us with high quality transit light curves that we will use in our
global analysis of \name{} b.

The data were taken with Spitzer's Infrared Array Camera
\citep[IRAC;][]{fazio04} as part of the Spitzer programs 542 (P.I. Des\'ert)
and 70049 (P.I. Deming). The observations feature three and fourteen transit
observations at 3.6 and 4.5 \micron{,} respectively. \cite{gillon14} also
presented six secondary eclipse observations of \name{} b that did not produce
any significant detection of the eclipse depth. As such, we elect to omit the
eclipse observations from our analysis and focus solely on the transit
observations.

The Spitzer photometry contains apparent variations in flux due to intra-pixel
sensitivity variations in the IRAC detectors. We corrected for this effect
using the pixel-level decorrelation (PLD) method \citep{deming15}, implementing
the procedure broadly as described by \cite{garhart20}. We performed aperture
photometry on all of the Spitzer data at both 3.6 and 4.5 \micron{,} using
eleven circular numerical apertures. The apertures had constant radii ranging
from tightly centered on the point spread function (PSF),
to much broader than the PSF (i.e. 1.6 to 3.5 pixels),
and were centered on the stellar image using a 2-D Gaussian fit. We
applied the PLD analysis to segments of the subsequent photometry, each segment
having a span of approximately $\pm 3$ hours and centered on the predicted times
of transit. In addition to fitting the photometry using the pixel coefficients
\citep{deming15}, we also fit a quadratic baseline in time. There are two
principal differences between our procedure and the analysis method described by
\cite{garhart20}. Firstly, we are fitting transits and not eclipses, so we
include quadratic limb darkening using the values calculated for the Spitzer
bandpasses \citep{claret12}. Secondly, at this phase we are interested only in
removing the instrumental systematic effects in the data, not yet fitting to the
transits. However, because the transits are intertwined with the instrument
effects, we include a preliminary transit fit simultaneously with removing the
intra-pixel systematic effects. We implement the transit portion of the fitting
process by fixing most orbital parameters to the values given by
\cite{charbonneau09}, with the exceptions being the time of mid-transit
and the transit depth, which are left as free parameters. The PLD code also
varies the coefficients of the quadratic ramp, and the pixel coefficients, whose
optimal values over the broad bandwidth are chosen using a Markov Chain Monte
Carlo method (see Sections 3.4 of \cite{garhart20} and Section 3.3 of
\cite{deming15}). We then remove the instrumental portion and retain the
detrended transit light curve for the more detailed fitting process described
in Section~\ref{sect:global}.

\subsection{Multi-color Photometric Monitoring with STELLA}
Numerous teams have presented data from long-term photometric monitoring
campaigns of \name{} and reported discrepant values of the stellar rotation
period ranging from 40 days to $\gtrsim 100$ days
\citep{berta11,narita13,nascimbeni15}. Here we retrieve on the precise long-term
photometry of \name{} from \cite{mallonn18}. These observations were obtained
between May 2012 and September 2016 with the WiFSIP
imager mounted on the robotic STELLA telescope at Teide Observatory in Tenerife,
Spain \citep{strassmeier04}. The observations were taken in the Johnson
$BV$ and Cousin $I$ bands. The light curves clearly reveal photometric
modulation by star spots with 5.3\% variability in the $BV$-bands and 1.3\% in
the $I$-band (Figure~\ref{fig:prot}). With these data, \cite{mallonn18} reported
a stellar rotation period of $125\pm 5$ days. This photometric variability
arising from magnetic activity will also induce an RV signal that we will model
in our joint transit plus RV analysis. We will use the STELLA photometry to
train our activity model as described in Section~\ref{sect:act}.

\section{Data Analysis \& Results} \label{sect:analysis}
\subsection{Training the RV Activity Model} \label{sect:act}
\name{} exhibits strong photometric variability at the level of 2-5\% in the
$BVI$ bands. Whether induced by dark star spots or bright photospheric faculae 
\citep{rackham17}, the magnetic activity will also be manifested in the RVs.
We opt to model the RV stellar activity with a semi-parametric Gaussian process
(GP), which models the temporal correlations in the RV data induced by
rotationally modulated stellar activity \citep[e.g.][]{haywood14}. Specifically,
we will model the RV activity as a quasi-periodic GP given that \name{} clearly
exhibits rotational modulation that is not strictly periodic due to the finite
lifetimes of active regions. The quasi-periodic covariance kernel is

\begin{equation}
  k_{ijp} = a_{\rm GP,b}^2 \exp{\left[ -\frac{(t_i-t_j)^2}{\lambda_{\rm GP} ^2} -\Gamma_{\rm GP}^2 \sin^2{\left( \frac{\pi |t_i-t_j|}{P_{\rm rot}} \right)} \right]}
    \label{eq:kernel}
\end{equation}

\noindent and is parameterized by the hyperparameters $a_{\rm GP,b}$, the
covariance amplitude in each of the $BVI$-bands (indexed by $b$), an
exponential timescale related to active region lifetimes $\lambda_{\rm GP}$, a
coherence parameter $\Gamma_{\rm GP}$, and the stellar rotation period \prot{.} To
first order, the origin of the photometric variability is common with the origin
of RV activity such that activity signals in each time series will show similar
temporal covariance. Consequently, we elect to train our GP activity model on
the photometry shown in Figure~\ref{fig:prot}. However, we 
note that chromospheric plages can induce a significant RV signal but with
little to no photometric signature, while near-polar active region
configurations can have the opposite effect of large photometric variations with
a small RV counterpart. As such, GP training on photometry in general may only
provide a partially complete picture of activity but any discrepancies for GJ
1214 are likely small due to the star's inactivity.

We sample the posterior of the logarithmic GP hyperparameters using the
\texttt{emcee} Markov Chain Monte Carlo (MCMC) code \citep{foremanmackey13} and
employ the \texttt{celerite} software package \citep{foremanmackey17} 
to construct the GP and evaluate its likelihood during the MCMC sampling. 
The mean GP model is depicted in Figure~\ref{fig:prot} in each passband along
with the associated $3\sigma$ uncertainty. The
corresponding hyperparameters are listed in Table~\ref{tab:hyperparam}.
We recover a photometric stellar rotation period of \prot{} $=$ \protval{}
days, which is consistent with the value reported by \cite{mallonn18}. In our
forthcoming transit plus RV modeling efforts, we will adopt the joint posterior
of $\{ \ln{\lambda_{\rm GP}}, \ln{\Gamma_{\rm GP}}, \ln{P_{\rm rot}} \}$ as
a prior on our RV activity model.

\input{training}

\subsection{Global Transit + RV Model} \label{sect:global}
Here we produce a global model of the available transit and RV observations.
Our primary goals are to refine the mass and orbital eccentricity of \name{} b
to aid in the scheduling of any future secondary eclipse observations.

We fit the detrended Spitzer transits in both IRAC channels with an analytical
transit model \citep{mandel02} computed using the \texttt{batman} software
\citep{kreidberg15}. Past TTVs searches have concluded that no significant TTV
signals persist in the system
\citep{carter11,harpsoe13,fraine13,gillon14}. As such, we assume a linear
ephemeris parameterized by the orbital period $P$ and the time of mid-transit
$T_0$. We also fit for the impact parameter $b$, orbital eccentricity $e$, and
the argument of periastron $\omega$. We mitigate the Lucy-Sweeney bias towards
non-zero eccentricities by sampling the parameters $h=\sqrt{e}\cos{\omega}$ and
$k=\sqrt{e}\sin{\omega}$ rather than sampling $e$ and $\omega$ directly
\citep{eastman13,lucy71}.

Furthermore, rather than sampling the scaled semimajor axis $a/R_s$ directly, we
sample the stellar mass $M_s$ and radius $R_s$ directly from their respective
measurement uncertainties, which we use together with $P$ to ensure
self-consistent values of $a/R_s$ with the aforementioned parameters.
Furthermore, by sampling $M_s$ and $R_s$, we can
compute the corresponding stellar density $\rho_s$, which is related to
the transit parameters $\{ P, e, \omega, a/R_s \}$ according to

\begin{equation}
  \frac{\rho_s}{\rho_{s,\text{transit}}} = \frac{(1-e^2)^{3/2}}{(1+e\sin{\omega})^3},
\end{equation}

\noindent \citep{moorhead11,dawson12} where

\begin{equation}
\rho_{s,\text{transit}} = \frac{3\pi}{G P^2} \left( \frac{a}{R_s} \right)^{3},
\end{equation}

\noindent \citep{seager03} where $G$ is the gravitational constant. These
expressions are intended to ensure consistency between the transit and
stellar, which when considered jointly with the RV measurements, 
provide the strongest possible constraints on the orbital eccentricity of
\name{} b.
In addition to the aforementioned transit parameters, we also fit separate sets
of the following wavelength-dependent parameters for the 3.6 and 4.5 \micron{}
channels: the planet-star ratio $r_p/R_s$, the quadratic limb-darkening
parameters $u_1$ and $u_2$, and the flux baseline $F_0$.

\begin{figure*}
  \centering
  \includegraphics[width=.9\hsize]{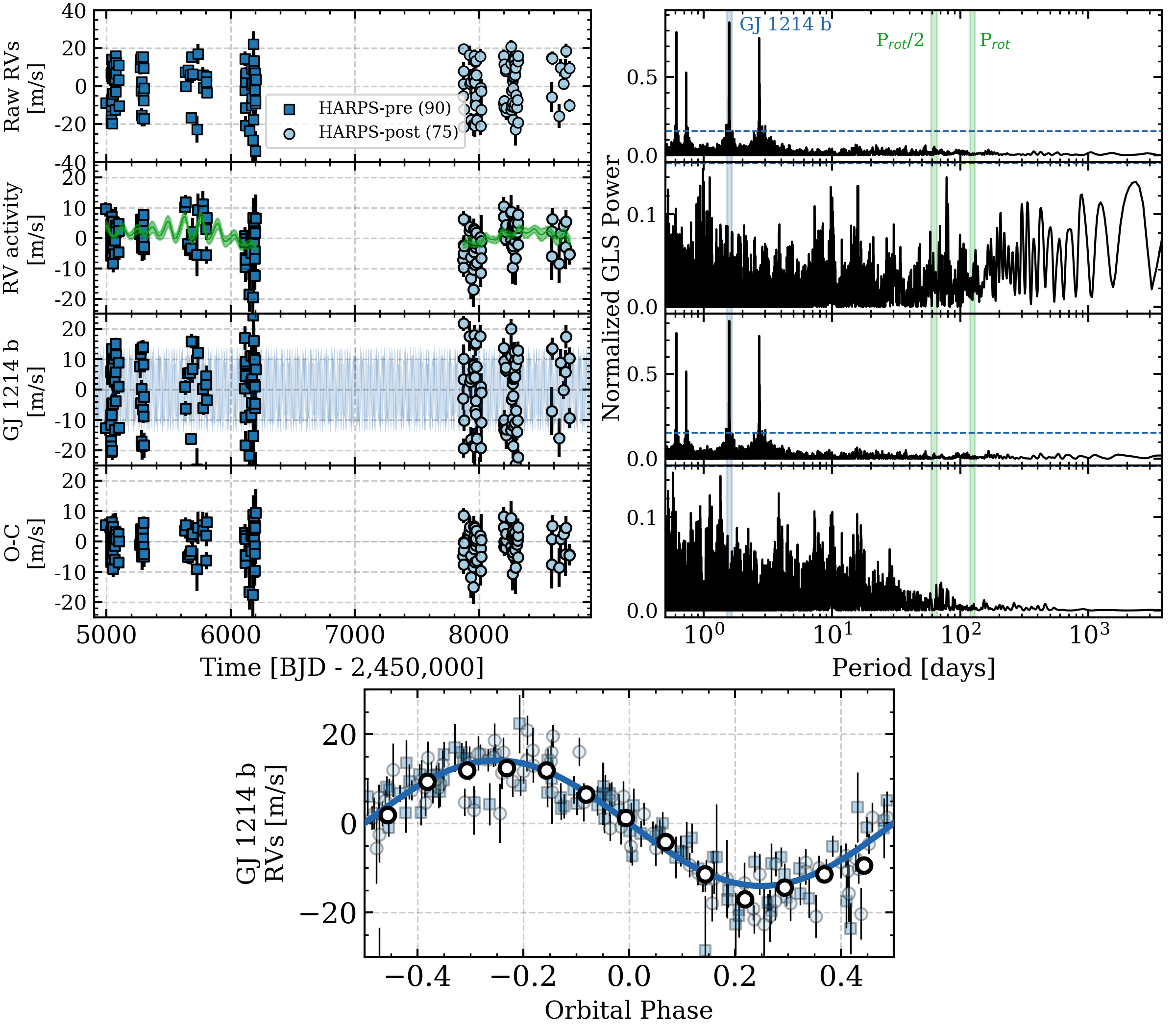}
  \caption{The pre (circles) and post (squares) fiber-upgrade HARPS RVs of GJ
    1214. In descending order, the left column depicts the RV time series of
    the raw RVs, the RV activity component, the \name{} b planetary component,
    and the RV residuals. The resulting residual rms values for the
      pre and post fiber-upgrade RVs are 6.2 and 4.7 \mps{,} respectively.
    The right column depicts the corresponding GLS
    periodogram. The vertical bands highlight the orbital period of \name{} b
    ($P=1.58$ days), the stellar rotation period (\prot{} $=124.7$ days) and
    its first harmonic ($P_{\rm rot}/2=62.4$ days). The dashed horizontal lines
    in each periodogram panel depict the 1\% FAP. Bottom panel: the
    activity-corrected RVs phase-folded to the orbital solution of \name{} b.}
  \label{fig:rv}
\end{figure*}

\begin{figure}
  \centering
  \includegraphics[width=\hsize]{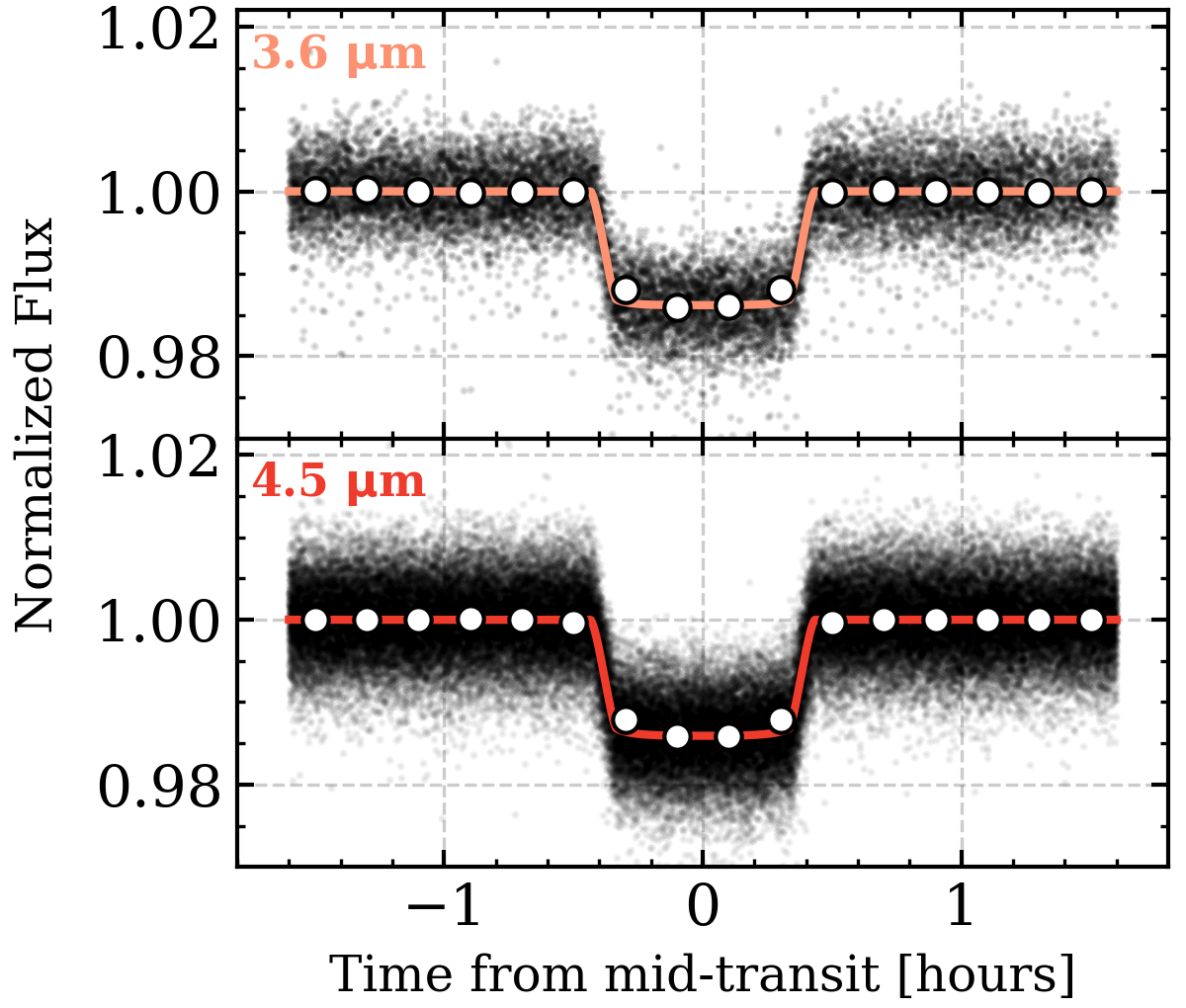}
  \caption{Phase-folded Spitzer/IRAC transit light curves of \name{} b. Upper
    panel: three 3.6 \micron{} transits. Lower panel: fourteen 4.5 \micron{}
    transits. Solid curves depict the MAP transit model
    in each IRAC channel. White circle markers depict the photometry binned
    for visualization purposes.}
  \label{fig:transit}
\end{figure}

We first correct the raw RVs for the deterministic secular acceleration of GJ
1214 (i.e. 0.3 m/s/yr). We then fit the RVs with a Keplarian orbital solution for
\name{} b plus an RV activity model trained on the STELLA photometry
(Section~\ref{sect:act}). In addition to the shared transit model parameters
$\{ P, T_0, h, k \}$, the Keplerian solution includes the RV semiamplitude $K$,
which we sample logarithmically in our MCMC. Our RV model also features velocity
offset $\gamma$ and logarithmic scalar jitter $\ln{s_{\rm RV}}$ parameters for the
subsets of our HARPS
RV time series corresponding to the pre and post fiber-upgrade. Similarly to the
GP training step, our RV activity model is described by the hyperparameters
$\{\ln{a_{\rm RV}},\ln{\lambda_{\rm GP}},\ln{\Gamma_{\rm GP}},\ln{P_{\rm rot}} \}$,
with the latter three being shared with our training model on the STELLA
photometry. Our full transit plus RV model consists of 24
parameters: $\{ M_s, R_s, P, T_0, b, h, k, (r_p/R_s)_{3.6}, u_{1,3.6}, u_{2,3.6}, F_{0,3.6},$
$(r_p/R_s)_{4.5}, u_{1,4.5}, u_{2,4.5}, F_{0,4.5}, \ln{K}, \gamma_{\rm pre}, \gamma_{\rm post}, \ln{s_{\rm pre}},$
$\ln{s_{\rm post}}, \ln{a_{\rm RV}}, \ln{\lambda_{\rm GP}}, \ln{\Gamma_{\rm GP}}, \ln{P_{\rm rot}} \}$.
Our adopted priors are outlined in Table~\ref{tab:results}.

Similarly to the training step, we sample the posterior of our transit plus RV
model parameters using the \texttt{emcee} MCMC code and use the \texttt{celerite}
package to construct the GP. We use the resulting maximum a-posteriori (MAP) GP
hyperparameters to construct the predictive GP distributions for the pre and
post fiber-upgrade time series. We adopt the mean functions of each GP posterior
as our best-fit activity models, which are highlighted in the second row of
Figure~\ref{fig:rv}. The MAP
values of the remaining parameters are reported in Table~\ref{tab:results}.
We use these parameters to compute the Keplerian orbital solution shown in
Figure~\ref{fig:rv} and the 3.6 and 4.5 \micron{} transit models
shown in Figure~\ref{fig:transit}.

\subsection{Resulting Physical Planetary Parameters} \label{sect:mr}
From our measured planetary parameters we refine the orbital and physical
parameters of \name{} b.
Similarly to \cite{gillon14}, our Spitzer/IRAC analysis reveals $r_p/R_s$ values
in the 3.6 and 4.5 \micron{} channels that are consistent within $1\sigma$.
We therefore average their marginalized posteriors and measure
a planetary radius of $r_p=$ \rplanet{} \Rearth{.}
Our analysis of 165 HARPS RV measurements provides the most detailed view of
the mass of \name{} b to date. We measure a 27$\sigma$ RV semiamplitude of
\Krv{} \mps{,} which corresponds to a $19\sigma$ planetary mass measurement of
$m_p=$ \mplanet{} \Mearth{.} 
We note that with our RV dataset, GJ 1214 b's mass measurement precision
is now limited by the uncertainty in the stellar mass with only a small
contribution from the RV semi-amplitude uncertainty. The same remains true
for the planet's radius whose measurement uncertainty is dominated by the
uncertainty in the stellar radius \cite{gillon14}. Taking $r_p$ and $m_p$
together, we recover a $13\sigma$ detection of the planet's bulk density:
\rhoplanet{} g cm$^{-3}$.

\begin{figure}
  \centering
  \includegraphics[width=\hsize]{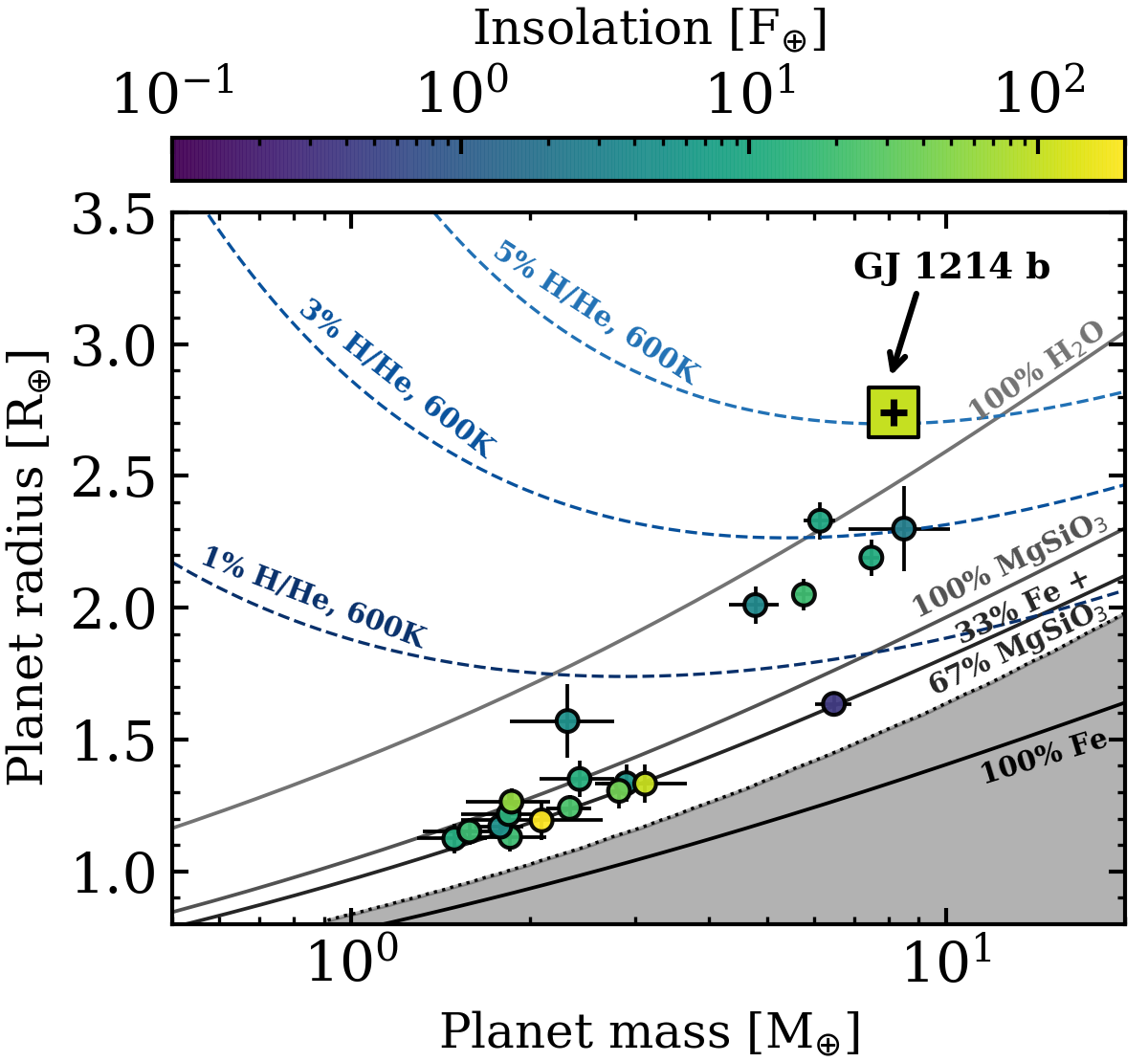}
  \caption{Mass–radius-insolation diagram for small planets with masses
    measured to $\leq 33$\% and orbiting mid-M dwarfs ($M_s \in [0.1,0.4]$
    \Msun{).} \name{} b is depicted by the large square marker. The solid curves
    depict internal structure models with mass fractions of 100\% water,
    100\% silicate rock, 33\% iron plus 67\% rock (i.e. Earth-like), and 100\%
    iron \citep{zeng13}. The dashed curves depict models of Earth-like cores
    enshrouded in solar metallicity H/He envelopes at the zero-albedo
    equilibrium temperature of GJ 1214 b (i.e. 600 K) and with various
    envelope mass fractions. The shaded region bounded by the
    dotted curve highlights the forbidden region according to models of
    maximum collisional mantle stripping by giant impacts \citep{marcus10}.}
  \label{fig:mr}
 \end{figure}

We include the refined mass and radius of \name{} b in the mass-radius diagram
of small planets orbiting mid-M dwarfs in Figure~\ref{fig:mr}.
When compared to two-layered, solid structure models \citep{zeng13}, we see
that \name{} b's low bulk density is clearly inconsistent with a purely solid
body and therefore requires an extended H/He envelope to explain its
mass and radius. With the inclusion of a gaseous envelope component, we are
unable to place meaningful constraints on the mass fraction of materials that
compose the planet's underlying structure. However, population synthesis models
of small close-in planets indicate that planets that form the sub-Neptune
peak in the M dwarf radius valley \citep{hirano18,cloutier20a} are likely
enveloped terrestrials\footnote{I.e. an Earth-like solid core surrounded by a
  H/He envelope with an envelope mass fraction of up to a few percent.} rather
than volatile-rich planets \citep{owen17,gupta19,rogers21}. If we assume that GJ
1214 b is in fact an enveloped terrestrial, we can calculate the envelope mass
fraction $X_{\rm env} \equiv m_{\rm env} / m_p$ required to explain its mass and
radius. We perform this calculation assuming that \name{} b consists of an
Earth-like core with a 33\% iron core mass fraction and hosts a H/He envelope
with solar-metallicity ($\mu=2.35$). The thermal structure of the gaseous
envelope is described by the semi-analytical radiative-convective model from
\citep{owen17}. We sample GJ 1214 b's joint $\{ m_p, r_p, T_{\rm eq} \}$ posterior
and calculate the core and envelope masses required to match the mass and radius
of \name{} b with an isothermal upper atmosphere at \teq{.} We measure an
envelope mass fraction of $X_{\rm env}=$ \Xenv{\%}. We note that the uncertainty
on $X_{\rm env}$ reported here is determined solely by the uncertainties in
the input parameters $\{ m_p, r_p, T_{\rm eq} \}$ and does not include
uncertainties in the physical model.

\subsection{Orbital Eccentricity Result} \label{sect:ecc}
Our joint transit plus RV analysis, together with a-priori constraints on the
stellar density, produces a strong constraint on \name{} b's orbital
eccentricity. The marginalized posterior on the orbital eccentricity
heavily favors a circular orbital solution. We find that $e<$ \ecc{} and
$<$ \eccnn{} at 95\% and 99\% confidence, respectively. Although a circular orbit
is expected given the planet's short circularization timescale of $<1$ Myr,
precise knowledge of the quantity $e\cos{\omega}$ has critical consequences for
scheduling secondary eclipse observations with facilities such as JWST. The
value of $e\cos{\omega}$ directly translates into an uncertainty window on the
eclipse center time according to

\begin{equation}
  \Delta t = \frac{2}{\pi}\,P\,e\cos{\omega}.
\end{equation}

\noindent Our stringent constraints on $e\cos{\omega}$ translate into 
an uncertainty on the center time of eclipse of 2.8 hours at 95\% confidence,
or approximately 3.2 times the eclipse duration of 53 minutes.

\section{Limits on Additional Planets} \label{sect:sens}
Continuous photometric monitoring of \name{} with Spitzer revealed no new
transiting planets larger than Mars and interior to the inner HZ at 20.9 days
\citep{gillon14}. Furthermore, TTV searches have not yielded any additional
planetary signals \citep{carter11,harpsoe13,fraine13,gillon14}. With our
intensive RV time series we are able to place meaningful constraints on
additional planets orbiting \name{,} which need not be in a transiting
configuration, nor locked into a mean-motion resonance with \name{} b where
the TTV sensitivity is maximized.
Here we derive limits on the existence of additional planets around
\name{} via a set of injection-recovery tests into our RV time series.

\subsection{Injection-Recovery Methodology} \label{sect:injrec}
We conduct a pair of Monte Carlo (MC) simulations to calculate the detection
sensitivity of planets as a function of planet mass and orbital period.
The MC simulations differ in their noisy RV time
series into which we will inject planetary signals.

\emph{MCI}: we sample the Gaussian distribution $\mathcal{N}(0,\sigma_{\rm eff})$
at the times of our HARPS pre and post-fiber upgrade window functions. The
effective RV noise $\sigma_{\rm eff}$ values are calculated from the quadrature
addition of the median RV uncertainties and the MAP scalar jitters from our RV
analysis ($\sigma_{\text{eff,pre}}=4.53$ \mps{,} $\sigma_{\text{eff,post}}=3.99$
\mps{).} However in practice, RV uncertainties are never purely
Gaussian, often as a result of variable observing conditions or
imperfect corrections of instrumental drifts. To mimic these
stochastic effects, we add a unique noise term to the RV dispersion by
sampling from $\mathcal{U}(-1,1)$ \mps{,} whose limits are
generally appropriate for HARPS observations.
The MCI simulations should be interpreted as a best case scenario in
that the noisy RVs are uncontaminated by temporally-correlated signals
from stellar activity, which are implicitly assumed to have been perfectly
mitigated.

\emph{MCII}: alternatively, we construct noisy RVs by subtracting
the MAP Keplerian signal of \name{} b from the pre and post fiber-upgrade RVs
(see second row in Figure~\ref{fig:rv}). In this way, signals from the window
function and stellar activity are preserved. The MCII simulations can be
interpreted as a worst case scenario wherein stellar activity is left
uncorrected and will contribute to the rms of the RVs. This effect will be
particularly detrimental to the recovery of injected planetary signals close
to \prot{} \citep{newton16,vanderburg16}. However undesirable the persistence of
stellar activity, and aliases, are for planet searches, these effects make the
MCII results a more realistic assessment of the detection sensitivity compared
to MCI.

In each of the $10^5$ MC realizations in MCI and MCII, we inject a single
synthetic Keplerian signal into the noisy HARPS RVs. We note that searching for
individual planets in our injection-recovery tests is a simplification as blind
searches for multiple RV planets are often conducted simultaneousy rather than
iteratively, such that the detection sensitivity is not independent of the
number of planetary signals.
The masses and periods of our synthetic planet population are sampled uniformly
in log-space between 0.1-20 \Mearth{} and 0.5-200 days, respectively. We select
these outer bounds with the intention to cover
the majority of the parameter space which is spanned by known planets around
mid-M dwarfs (see Section~\ref{sect:freq}). Our adopted mass range encapsulates
planets with masses smaller than that of Neptune (i.e. 17 \Mearth{)}.
The mass of \name{} is sampled from
its posterior $\mathcal{N}(0.178,0.010)$ \Msun{.} Due to the transiting nature
of \name{} b and the characteristically low dispersion in mutual inclinations
among Kepler multi-planet systems with $a/R_s>5$ \citep{dai18}\footnote{Note
  that for \name{} b, $a/R_s=14.85\pm 0.16$.}, we focus on orbital inclinations
that are nearly co-planar with \name{} b ($i_b=88.7\pm 0.1^{\circ}$) by sampling
inclinations from $\mathcal{N}(i_b,\sigma_i)$, where $\sigma_i=2^{\circ}$
\citep{ballard16,dai18}. Planets' orbital phases are sampled from
$\mathcal{U}(-\pi,\pi)$ and all orbits are assumed to be circular.

We inject the synthetic planetary signal in each realization and attempt
to recover the injected signal based on two criteria in MCI plus one additional
criterion in MCII. Because the search
for non-transiting planets lacks the luxury of a-priori knowledge of the
planet's ephemeris, a non-transiting planet's signal in an RV time series must
emerge as a prominent periodicity in the data. As such, our first criterion for
the successful recovery of an injected planet is that the maximum GLS periodogram
power within 10\% of the injected period must exceed a FAP of 1\%. However,
sporadic peaks in the GLS periodogram can masquerade as a planetary signal.
We safeguard against these false positives by insisting that a
six-parameter Keplerian model at the GLS peak's orbital period must be strongly
favored over the null hypothesis (i.e. a flat line with a velocity offset). For
this model comparison, we calculate the Bayesian Information Criteria (BIC) for
the Keplerian model and for the null model.
The BIC $=-2\ln{\mathcal{L}} + \nu\ln{N}$, where $\mathcal{L}$ is
the likelihood of the RV data given the model, $\nu$ is the number of model
parameters ($\nu=6$ or 1 for the Keplerian and null
models, respectively), and $N=165$ is the number of RV measurements. Taken
together, we claim the successful recovery of an injected planet in MCI if 1)
the GLS power of the injected
periodic signal has FAP $<1$\%, and 2) the BIC of the null hypothesis is
greater than ten times the BIC of the Keplerian model.

\begin{figure*}
  \centering
  \includegraphics[width=.9\hsize]{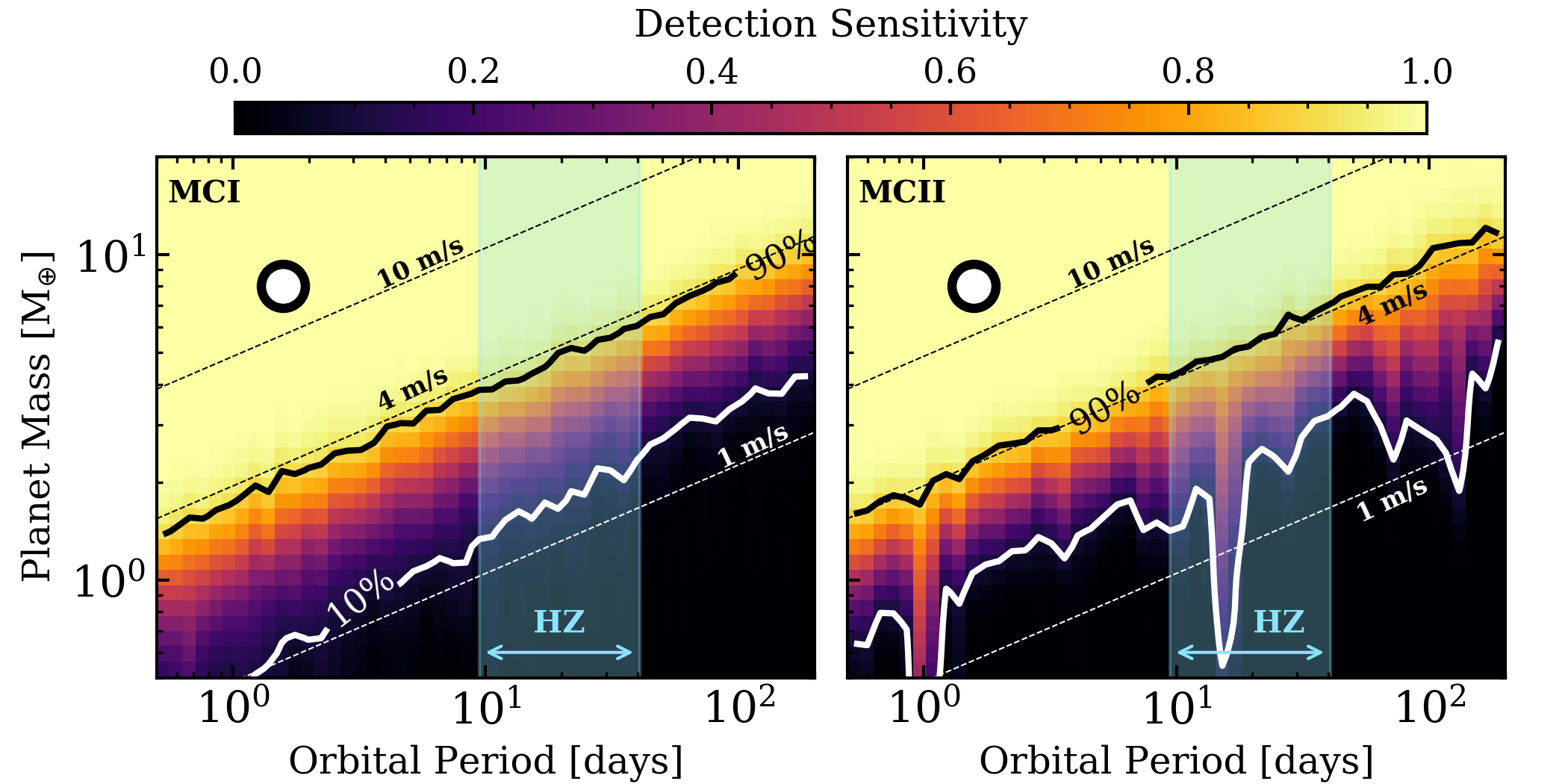}
  \caption{RV detection sensitivity to planets orbiting \name{} as a function
    of planet mass and orbital period for two noise models. Left panel: the
    results of the MCI simulation, which should be interpreted as a best case
    scenario. Right panel: the results of the MCII simulation, which should be
    interpreted as the worst case scenario (see Section~\ref{sect:injrec} for
    details). The solid contours highlight the 10\% and 90\% sensitivity levels.
    The thin dashed
    lines represent lines of constant semiamplitude with illustrative values
    equal to 1, 4, and 10 \mps{.} The circle markers highlight \name{} b. The
    vertical shaded regions span the habitable zone of \name{.}}
  \label{fig:sens}
\end{figure*}

The noisy RVs in MCI do not contain any temporally correlated signals such
that there are no significant peaks in the GLS periodogram of the noisy RVs.
However, the MCII noisy RVs retain the stellar activity signals which can
produce GLS periodogram peaks at \prot{,} one of its low order harmonics, or
at spurious periods unrelated to stellar rotation \citep{nava20}.
We therefore apply a third condition to the successful recovery of an injected
Keplerian signal in MCII. Namely, the largest peak in the GLS periodogram of
the noisy RVs within 10\% of the injected period must have FAP $>1$\%. This
is to ensure that the injected signal is not confused with a residual signal
or an alias close to the its orbital frequency. 
Alternatively, if the BIC of the null hypothesis is
greater than $10^3$ times the BIC of the Keplerian model, then we continue to
accept the injected signal as a detection because such RV signals have very high
S/N and should not be rejected by relatively minor aliases present in the window
function.

Due to the exhaustive nature of computing the GLS periodogram and its
corresponding FAP in $2\times 10^5$ realizations via bootstrapping with
replacement, we
adopt an analytical framework to approximate the FAP as a function of the
normalized GLS power $p$ \citep{cumming04}. For a time series containing $N$
measurements and spanning $T$ days, the FAP is well-approximated by

\begin{equation}
  \text{FAP}(p_0) = 1 - [ 1 - \text{Pr}(p > p_0)]^M,
  \label{eq:gls}
\end{equation}

\noindent where $M=\Delta f / \delta f$ is the number of sampled independent
frequencies and is calculated from the ratio of the frequency range
$\Delta f = f_2 - f_1$ and the frequency resolution $\delta f=1/T$. For a GLS
periodogram normalized by the sample variance, such that the GLS power is
bounded between 0 and 1 \citep{zechmeister09}, the probability term in
Eq.~\ref{eq:gls} becomes
$\text{Pr}(p > p_0) = (1-p_0)^{(N-3)/2}$ \citep{cumming99}, where $p_0$ is the
power whose FAP we are interested in. Adopting this analytical formalism
significantly decreases the computation time of the MC simulations and
enables the mass-period space to be sampled more efficiently.

\subsection{RV Detection Sensitivity} \label{subsect:sens}
We calculate the RV detection sensitivity in each MC simulation as
the ratio of the number of recovered planets to the number of injected planets
as a function of mass and period. Our results are shown in
Figure~\ref{fig:sens}.

The results from MCI indicate that the sensitivity is a smooth function of RV
semiamplitude. Injected planetary signals with $K\gtrsim 4$ \mps{} are
detected $>90$\% of the time. The sensitivity contours derived from MCII are
broadly similar with the 90\% contour being well-aligned with $K=4$ \mps{} as in
MCI. However in MCII, residual low FAP frequencies in the HARPS RVs after the
removal of \name{} b disrupt the smoothness of the map. In particular,
residual GLS peaks introduce structure in the sensitivity map, which can in
principle inhibit the detection of planetary signals whose orbital periods are
close to low FAP peaks (or aliases thereof) in the window function.
However, the persistent periodicities seen in the GJ 1214 RV residuals at
1, 16, and 125 days have FAPs $>1$\% such that they are not confused with
injected signals with similar orbital periods.
But because these periodicities do exhibit a
slightly enhanced GLS power, the GLS power of injected planets near these
frequencies is artificialy enhanced. This effect manifests
itself as a moderate improvement to the sensitivity in the second panel of
Figure~\ref{fig:sens}.  We caution that these enhancements are an undesirable
outcome of a periodogram-based recovery process and should not be taken at
face value in a planet search.

The residual signals at 1 and 125 can be easily explained by the daily alias
and the rotation period of GJ 1214,
respectively. However, the signal at approximately 16 days is not trivially
explained by a stellar rotation harmonic or by an alias from the HARPS window
function. The FAP of the 16-day signal in the MCII RV residuals (i.e. without
removing a GP) exceeds 1\% such that the evidence for a periodic signal at 16
days is marginal. 
Furthermore, the GLS periodogram of the RV residuals following the removal of
GJ 1214 b and the GP activity model does not show any significant power at 16
days (see the fourth row in Figure~\ref{fig:rv}). As a
check, we conducted a separate RV analysis that included a second Keplerian
signal with a narrow uniform prior on its orbital period centered on 16
days. The resulting marginalized posterior on the Keplerian semiamplitude is
consistent with zero and has an upper limit of 2.45 \mps{} at 95\% confidence
(i.e. corresponding to \msini{} $<3.03$ \Mearth{)}. Together these tests
indicate that there is no strong evidence for a second planet near 16 days
but continued RV follow-up will help to establish the nature of this signal.

\begin{figure}
  \centering
  \includegraphics[width=.9\hsize]{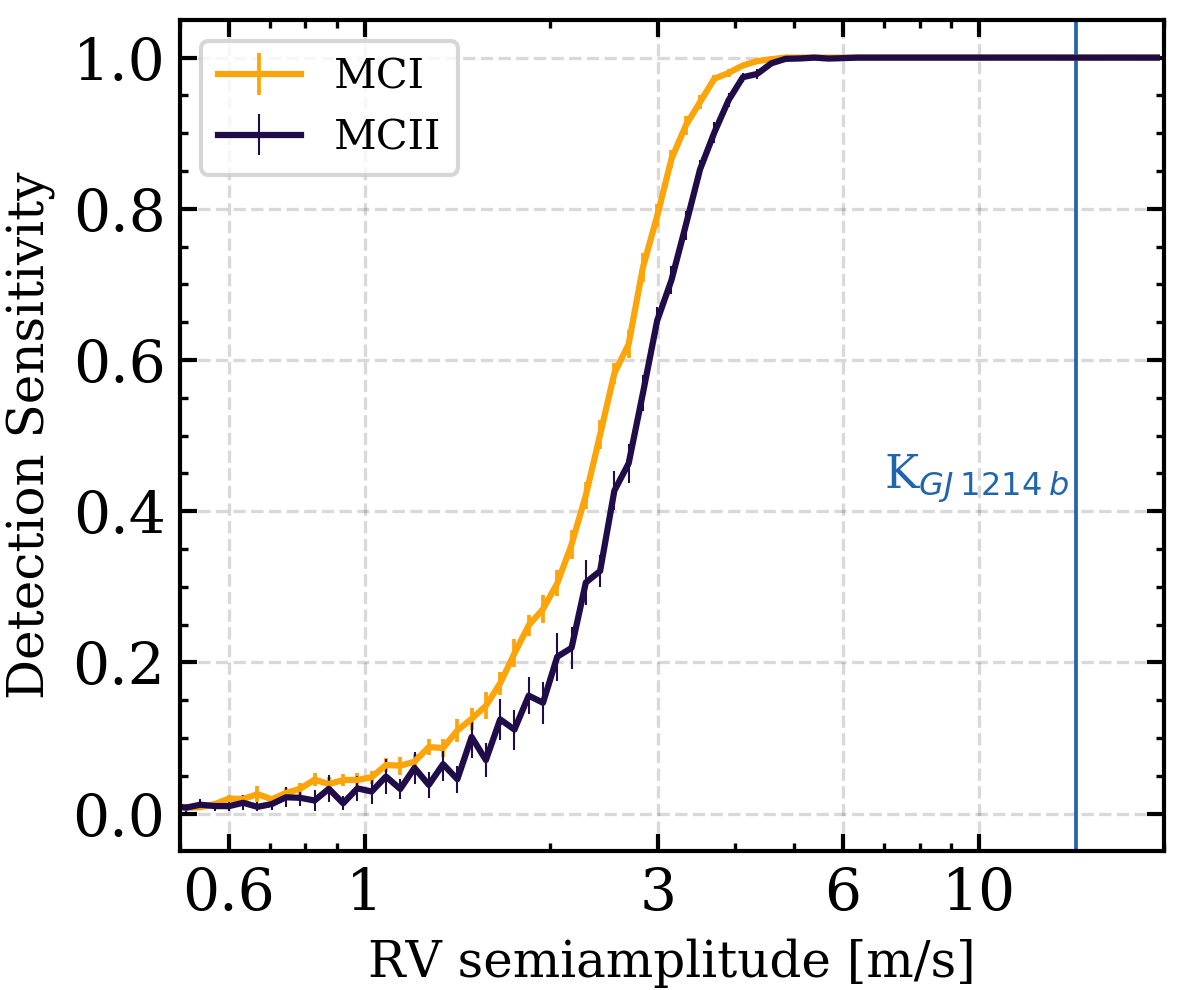}
  \caption{RV detection sensitivity to planets orbiting \name{} as a function
    of RV semiamplitude for two noise models. The gold and blue curves depict
    the RV detection sensitivity from the MCI and MCII simulations,
    respectively. The errorbars illustrate the error on the mean detection
    sensitivity in each semiamplitude bin. The vertical solid line highlights
    the semiamplitude of \name{} b.}
  \label{fig:sensK}
\end{figure}

Modulo the detailed structure in the MCII detection sensitivity, the results
are broadly consistent with MCI. The similarity between each MC simulation's
detection sensitivity is depicted in Figure~\ref{fig:sensK} as a function of
$K$. The two MC simulation curves are in close agreement for all values of $K$
with the MCII sensitivity exhibiting a slight detriment relative to MCI
up to $K \geq 5$ \mps{,} at which point all planetary signals within 200 days
are detectable. Focusing on the more realistic results from MCII and taking the
90\% contour to mark the boundary between detection and non-detection,
we are able to rule out additional planets around \name{} above 4 \Mearth{} and
within 10 days. By adopting the empirical recent Venus and early Mars limits of
the inner and outer habitable zone (HZ) from \cite{kopparapu13}, we can rule out
planets down to 5 \Mearth{} within the HZ. The most massive Super-Earths
have masses $\lesssim 7$ \Mearth{} (LHS 1140 b; \citealt{ment19}, TOI-1235 b;
\citealt{cloutier20c}), such that we are only able to rule out the most massive
HZ rocky planets.

Within our sensitivity limits, \name{} continues to be a system with only one
known planet. In the next section, we will argue that this result is somewhat
surprising. We will show that multi-planet systems around mid-M dwarfs are
common and that \name{} belongs to a small subset of mid-M dwarf systems
that appear to be singles given the current RV data available.

\section{The Frequency of Multi-Planet Systems Around Mid-M Dwarfs}
\label{sect:freq}
The Kepler transit survey revealed that multi-planet systems around
early M dwarfs are ubiquitous
\citep[e.g.][]{dressing15a,ballard16,gaidos16,hsu20}. Studies focused on later
type M dwarfs have shown preliminary evidence of a trend of increasing
small planet multiplicity with decreasing stellar mass. Although the current
inferences are largely limited by poor counting statistics around mid-to-late
M dwarfs \citep{muirhead15,hardegree19}. If it is true that mid-M dwarf
planetary systems frequently host multiple planets, then the confirmation of
even a single transiting planet around a mid-M dwarf would establish a strong
prior on the existence of additional planets. Indeed, a number of apparently
single transiting mid-M dwarf systems have revealed additional planets in
RV follow-up campaigns (GJ 357c,d; \citealt{luque19a}, GJ 1132c;
\citealt{bonfils18b}, GJ 3473c; \citealt{kemmer20}, LHS 1140c;
\citealt{ment19}).
Here we aim to quantify this phenomenon by calculating the frequency of mid-M
dwarf planetary systems that host multiple planets. 

\subsection{Calculating the Frequency of Mid-M Dwarf Planetary Systems with Multiple Planets}
Let $f$ denote the fraction of mid-M dwarf planetary systems that host
two or more planets over some range of planet masses and orbital periods.
That is, among the sample of mid-M dwarfs that host at least one known
planet, $f$ denotes the fraction of those planetary systems that host
multiple planets. To calculate $f$, we will consider $N$ mid-M
dwarfs that host at least one known planet (see Section~\ref{subsect:stars}),
which we index by $i=1,\dots,N$. For each mid-M dwarf in our sample, we define
$p_i \equiv$ the RV detection sensitivity to additional planets over the range of
planet masses $\in [m_{p,\rm min},m_{p,\rm max}]$ and orbital periods
$\in [P_{\rm min},P_{\rm max}]$. We will compute the $p_i$ values in
Section~\ref{subsect:pi}.  Because each planetary system
contains at least one confirmed planet, the probability of detecting an
additional planet proceeds as a Bernoulli process with probability $fp_i$. We
denote the results of the $N$ Bernoulli experiments by the set of binary values
$\{ k_i \}$ wherein $k_i=1$ if the $i^{\rm th}$ system contains multiple planets,
with at least one planet having
$m_p \in [m_{p,\rm min},m_{p,\rm max}]$ and $P \in [P_{\rm min},P_{\rm max}]$.
Otherwise, $k_i=0$. For example, in Section~\ref{subsect:sens} we showed
that GJ 1214 appears to be a single-planet system such that $k_{\rm GJ 1214}=0$,
whereas for the two-planet system LHS 1140, $k_{\rm LHS 1140}=1$.
The likelihood of measuring $\{ k_i \}$ from $N$ systems
is given by

\begin{equation}
  \mathcal{L}(\{ k_i \} | f, \{p_i\}) = \prod_{i=1}^{N}\: (fp_i)^{k_i}\: (1-fp_i)^{1-k_i}.
\end{equation}

\noindent Assuming that all values of $f$ are equally probable (i.e.
$\text{Pr}(f)=1$), the normalized posterior of $f$ follows from Bayes theorem:

\begin{equation}
  \text{Pr}(f|\{k_i \},\{p_i \}) = \frac{\mathcal{L}(\{ k_i \} | f, \{p_i \})}{\int_{0}^{1}\mathcal{L}(\{ k_i \} | f, \{p_i \}) df},
  \label{eq:f}
\end{equation}

\noindent which we can use to calculate $f$ and its corresponding confidence
intervals.

\subsubsection{Sample of Mid-M Dwarf Systems} \label{subsect:stars}
We proceed with measuring $f$ by compiling the set of mid-M dwarfs with

\begin{enumerate}
\item $M_s \in [0.1,0.4]$ \Msun{,} and
\item that host at least one confirmed planet with an RV mass measurement.
\end{enumerate}

\noindent Because our second condition requires systems to contain at least
one confirmed planet with an RV mass measurement, we only consider
transiting planetary systems because the absolute masses of non-transiting
planets uncovered in RV datasets are unmeasurable.
The existence of a mass detection also  (largely) implies that sufficient RV
follow-up was conducted and a robust RV detection
sensitivity has been achieved. This will enable meaningful constraints
on $f$ to be produced.
We note that our second criterion excludes systems with TTV masses only. This is
desirable given the difficulty in accessing the TTV sensitivity to additional
planets; a function which itself is not smooth with orbital separation due to
the dependence of the TTV amplitude on the proximity of planet pairs to
low-order mean motion resonances.
We identify \Ntransit{} transiting planetary systems based on our criteria and
list them in Table~\ref{tab:midtolate}.

\input{midtolatetransiting}

One notable system is excluded from our sample on
the basis of having TTV masses only without substantial RV follow-up
\citep[K2-146;][]{hamann19}. Other notable
exclusions include the nearby system LHS 3844 \citep{vanderspek19} and the
multi-transiting system around LP 791-18 \citep{crossfield19}, both of which
lack published RV time series.

\subsubsection{RV Detection Sensitivity} \label{subsect:pi}
We compute each system's RV detection sensitivity $p_i$ identically to the
procedure described in Section~\ref{sect:injrec} as applied to our \name{} RV
data (i.e. via injection-recovery tests in the RV time
series). We retrieve each system's RV time series from its literature source
listed in Table~\ref{tab:midtolate}. For the MCI simulations, we construct the
noisy RVs from the quadrature addition of the median RV uncertainty and
the scalar jitter for each unique spectrograph. In the MCII simulations, we
subtract the orbtial signal for each known planet using the Keplerian parameters
reported in the reference. The planet injections and subsequent recoveries
proceed identically as with \name{.} The resulting sensitivity maps for each
mid-M dwarf are displayed in Appendix~\ref{app:sens}.

\subsubsection{The Frequency of Multi-Planet Systems Around Mid-M Dwarfs}
\label{subsect:freq}
\begin{figure}
  \centering
  \includegraphics[width=\hsize]{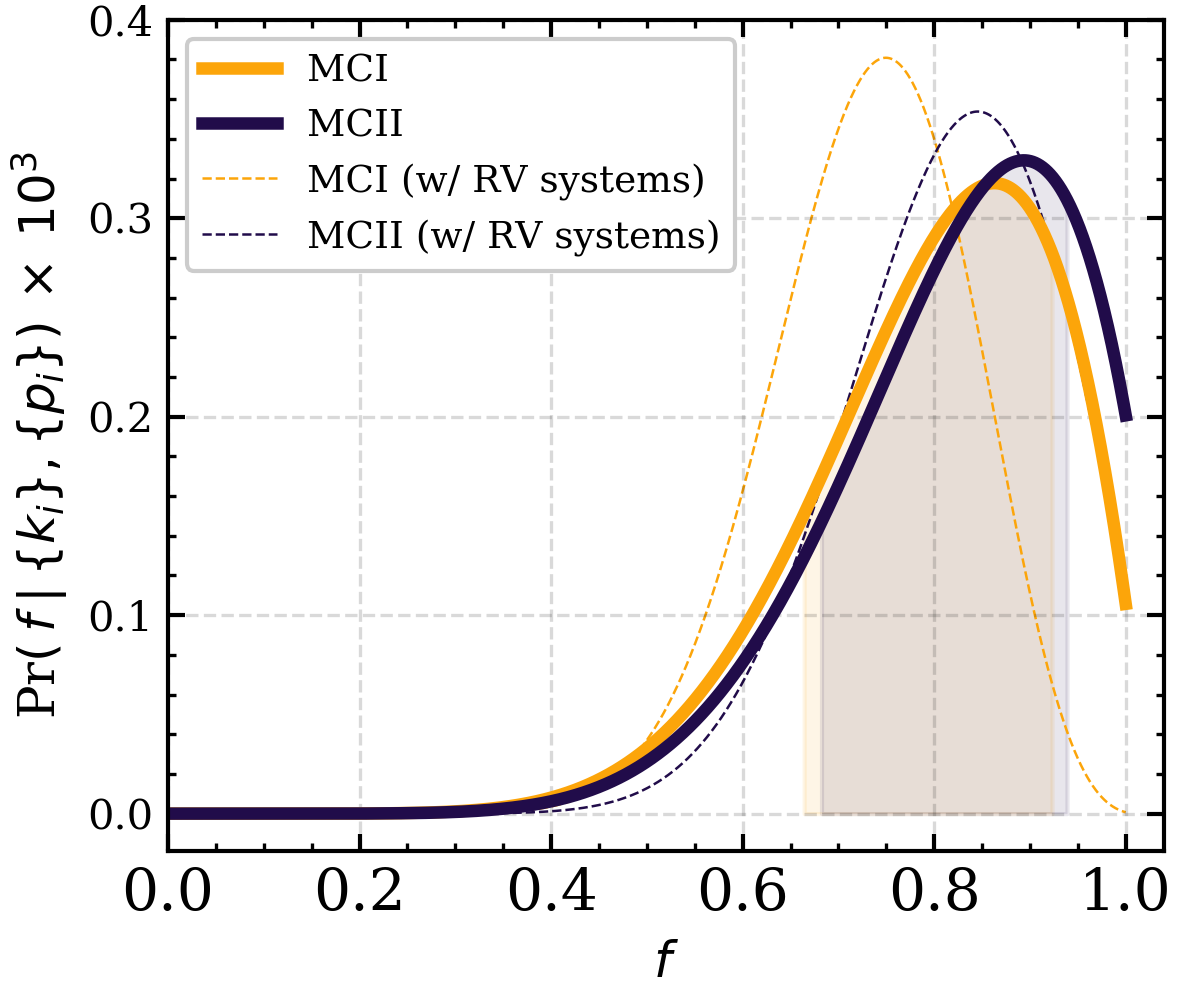}
  \caption{The posterior probability of $f$: the fraction of mid-M dwarf
    planetary systems that
    host multiple planets with masses $m_p \in [1,10]$ \Mearth{}
    and orbital periods $P \in [0.5,50]$ days. The analytical $f$ posterior
    is calculated using Eq.~\ref{eq:f} given the mid-M dwarfs with
    multiple planets $\{ k_i \}$ and their respective RV detection sensitivities
    $\{ p_i \}$. The solid gold curve depicts the result using the detection
    sensitivities from MCI whereas the solid blue curve depicts the result
    from MCII (see Section~\ref{sect:injrec}).
    As a check, the dashed curves depict the $f$ posterior from
    each MC when RV planetary systems are included in the calculation (see
    Section~\ref{sect:rvf}).}
  \label{fig:fpostv2}
\end{figure}

After tabulating $\{ k_i \}$ and calculating $\{ p_i \}$ among the mid-M
dwarfs in our sample, we compute the posterior of $f$ using Eq.~\ref{eq:f}.
Due to the relatively small number of mid-M dwarf systems with at least one
known transiting planet, we consider a wide range of planetary parameters
bounded by $m_p \in [1,10]$ \Mearth{} and $P \in [0.5,50]$ days.
However, because the typical RV detection sensitivity varies widely
across such a wide range of parameters (c.f. Figure~\ref{fig:sens}), our
results as discussed below only apply under the assumption that planets
are distributed uniformly as a function of $\log{m_p}$ and $\log{P}$, both of
which are implicitly assumed based on our sampling of injected planets. 
Of the \Ntransit{} systems in our sample, eight host a multi-planet system with
at least one planet within the aforementioned mass and period bounds. The
resulting $f$ posteriors from MCI and MCII are depicted in
Figure~\ref{fig:fpostv2}. The MAP
values of $f$, along with confidence intervals bounded by the 16$^{\rm th}$ and
84$^{\rm th}$ percentiles are \ftransitclean{} and \ftransitminuskep{} for MCI
and MCII, respectively. The 95\% lower limits on $f$ are $>57$\% and $>59$\% for
MCI and MCII, respectively. The consistency of $f$ between both MC simulations
highlights the insensitivity of our results to the exact
method of constructing the noisy RVs.

The consensus from our calculations is that the most likely fraction of mid-M
dwarf planetary systems that host multiple planets less massive than 10
\Mearth{} and within 50 days is approximately 90\%. To rephrase,
the discovery of a single planet within these bounds around a mid-M dwarf
establishes a strong prior that favors the existence of at least one additional
planet in the system.

\subsubsection{Including RV Systems When Calculating $f$} \label{sect:rvf}
\begin{figure}
  \centering
  \includegraphics[width=\hsize]{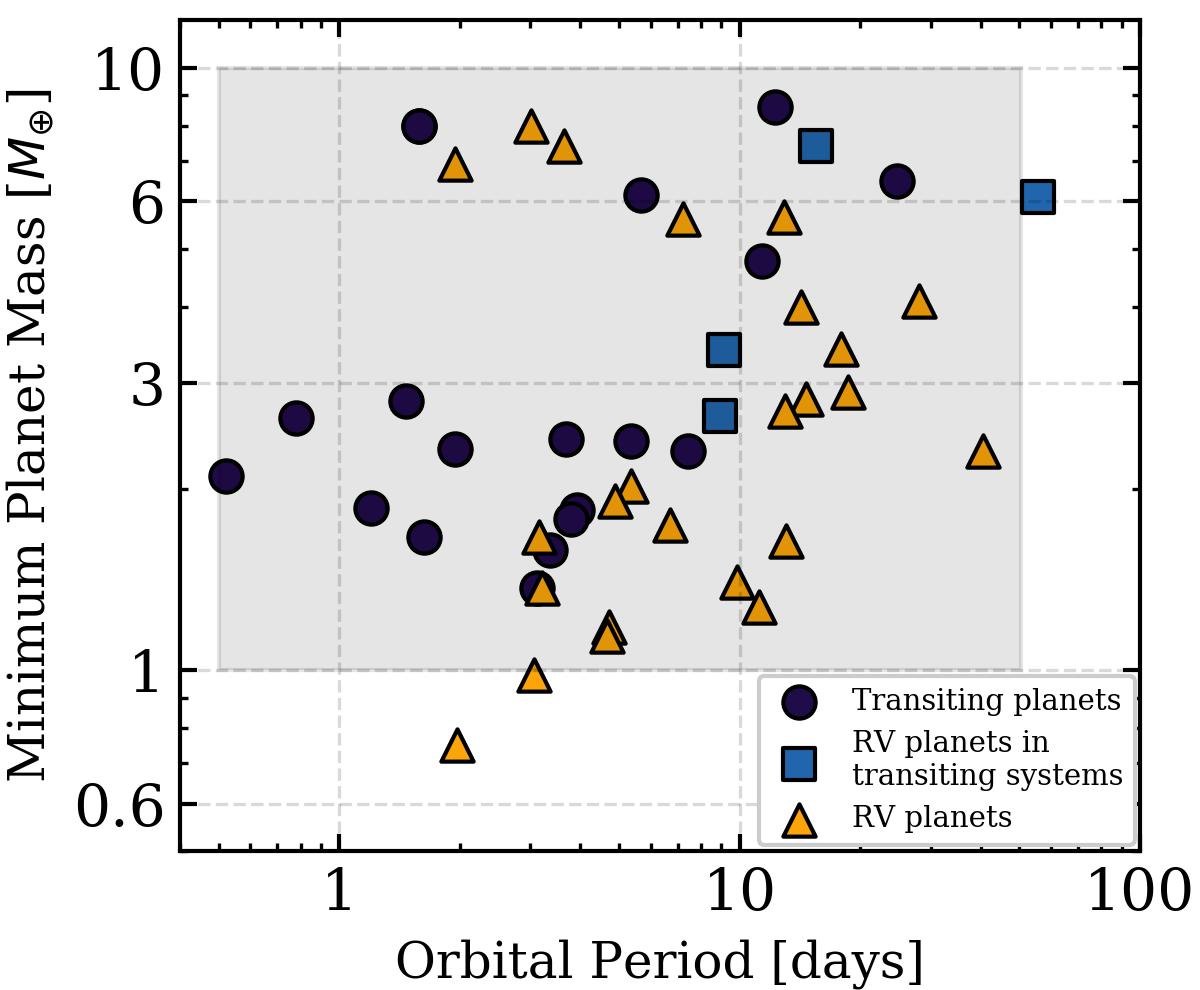}
  \caption{Minimum masses and orbital periods of planets orbiting mid-M dwarfs
    and with an RV mass detection. The circles depict 19 transiting planets in
    \Ntransit{} systems. The squares depict four RV planets in three transiting
    planetary systems. The triangles depict the 31 RV planets in \Nrv{}
    systems. The shaded rectangle outlines the range of planetary parameters
    over which we calculate $f$ in this work.}
  \label{fig:Pmp}
\end{figure}

Our previous calculation of $f$ focused solely on transiting planets whose
absolute masses are measured with an RV detection. Here we relax this
requirement by equating minimum planet mass \msini{} to planet mass, which
allows us to include \Nrv{} additional mid-M dwarf planetary systems
(Table~\ref{tab:midtolateRV}). We construct our RV planet sample by
retrieving planets with \msini{} $\in [1,10]$ \Mearth{,} $P \in [0.5,50]$
days, and whose dispositions have not been directly contested
in the literature. Our sample of RV
planets is compared to the transiting planet sample in $P$-$m_p$ space in
Figure~\ref{fig:Pmp}. Most of the RV planets have minimum masses and periods
that are distinct from the transiting planet sample. RV planets often have
longer orbital periods because of the geometric selection effect of decreasing
transit probability with increasing orbital period. Many of the RV planets
are also less massive than the transiting planets
because their host stars tend to be brighter, which often
results in more intensive RV follow-up and with higher S/N measurements, both of
which facilitate the detection of low amplitude signals.

\input{midtolaterv}

We wish to include the set of confirmed RV planets into our calculation of $f$.
We proceed by computing each system's detection sensitivity $\{ p_i \}$ via
injection-recovery in a manner similar to the transiting planetary systems. The
only difference being that we modify the distribution from which the injected
planet's inclinations are drawn. Because the orbital inclinations $i$ of RV
planets are unconstrained, we draw $i$ from the distribution that is uniform
in $\cos{i}$: i.e. $i$ drawn from $\arccos{\mathcal{U}(-1,1)}$. In general, this
geometric effect is detrimental to the RV detection sensitivity because it
allows for the possibility of nearly face-on orbits for which the RV
semiamplitude goes to zero given its linear dependence on $\sin{i}$. The
resulting sensitivity maps for each mid-M dwarf in our RV sample are included
in Appendix~\ref{app:sens}.

By combining the \Nrv{} RV systems with the \Ntransit{} transiting systems, we
expand the sets $\{ k_i \}$ and $\{ p_i \}$.
We note that tentative evidence for additional planet candidates
orbiting Proxima Centauri at 5.15 days \citep{suarez20} and at 5.2 years
\citep{damasso20} have been independently reported. However, due to the
present uncertainties regarding the nature of these signals, here we will
treat Proxima Centauri as a single-planet system (i.e. $k_{\rm Prox}=0$).
We note that the resulting value of $f$ is largely insensitive to any one
system and that if instead we set $k_{\rm Prox}=1$, the resulting value of
$f$ will only increase by 3\%. With $8/15=53$\% of RV
systems hosting multiple planets over the range of interest, the sample of RV
systems exhibits a slightly lower fraction of known multi-planet systems
compared to the transiting systems ($8/12=68$\%). However despite having
fractionally fewer multi-planet systems and systematically larger detection
sensitivities, we find that the $f$ values obtained when RV systems are included
in the calculation are consistent with those obtained from transiting systems
only (Figure~\ref{fig:fpostv2}). Explicitly, including RV planets in MCI we
measure \frvclean{,} compared to \ftransitclean{} with transiting systems alone.
Similarly in MCII, we measure \frvminuskep{,} compared to \ftransitminuskep{}
with transiting systems alone. Therefore the following statement continues to
be robust: the existence of one known planet orbiting a mid-M dwarf establishes
a strong prior that favors the existence of at least one additional planet in
the system.

\subsection{Is the Architecture of the \name{} System Unique Among Mid-M Dwarfs?}
In Section~\ref{subsect:freq} we calculated that \ftransitminuskep{} of
mid-M dwarf planetary systems host multiple planets when at least one planet has
$m_p \in [1,10]$ \Mearth{} and $P\in [0.5,50]$ days. Given that \name{} hosts
one such planet, and that multi-planet mid-M dwarf systems are extremely common,
it may appear contradictory that \name{} does not show evidence for additional
planets (Section~\ref{subsect:sens}). The cause of this apparent discrepancy
must be due to either the \name{} system being a bona-fide single-planet system,
or that additional planets are present but are undetected due to our
imperfect RV sensitivity.

Given \name{'s} detection sensitivity over the parameter range of interest
(Figure~\ref{fig:sens}), we find the probability that \name{} is a single-planet
system to be \fonep{\%}. This value is consistent with 50\% such that
there is no strong evidence that \name{} is a bona-fide single-planet
system. For example, if intensive RV characterization were to continue to
achieve a detection sensitivity that is on par with that of the mid-M dwarf
transiting system with the highest RV sensitivity to-date
\citep[$p_{\rm GJ 486}=90$\%;][]{trifonov21}, then we would
measure the probability that \name{} is a single-planet system to be
$15^{+21}_{-1}$\% (if it remained a single-planet system).
We estimate that reaching this goal would require an additional 120 hours of
observing time with HARPS \citep{cloutier18}. But using an alternative
spectrograph like MAROON-X, which is optimized for M dwarfs \citep{seifahrt18},
the observing time may be signficiantly reduced to about 15
hours.\footnote{\href{http://www.maroonx.science}{MAROON-X ETC}.} We emphasize
however that this calculation is approximate as it does not consider telescope
scheduling, overheads, and the unknown effects of 
effects of stellar activity at red optical wavelengths.

\begin{figure*}
  \centering
  \includegraphics[width=.9\hsize]{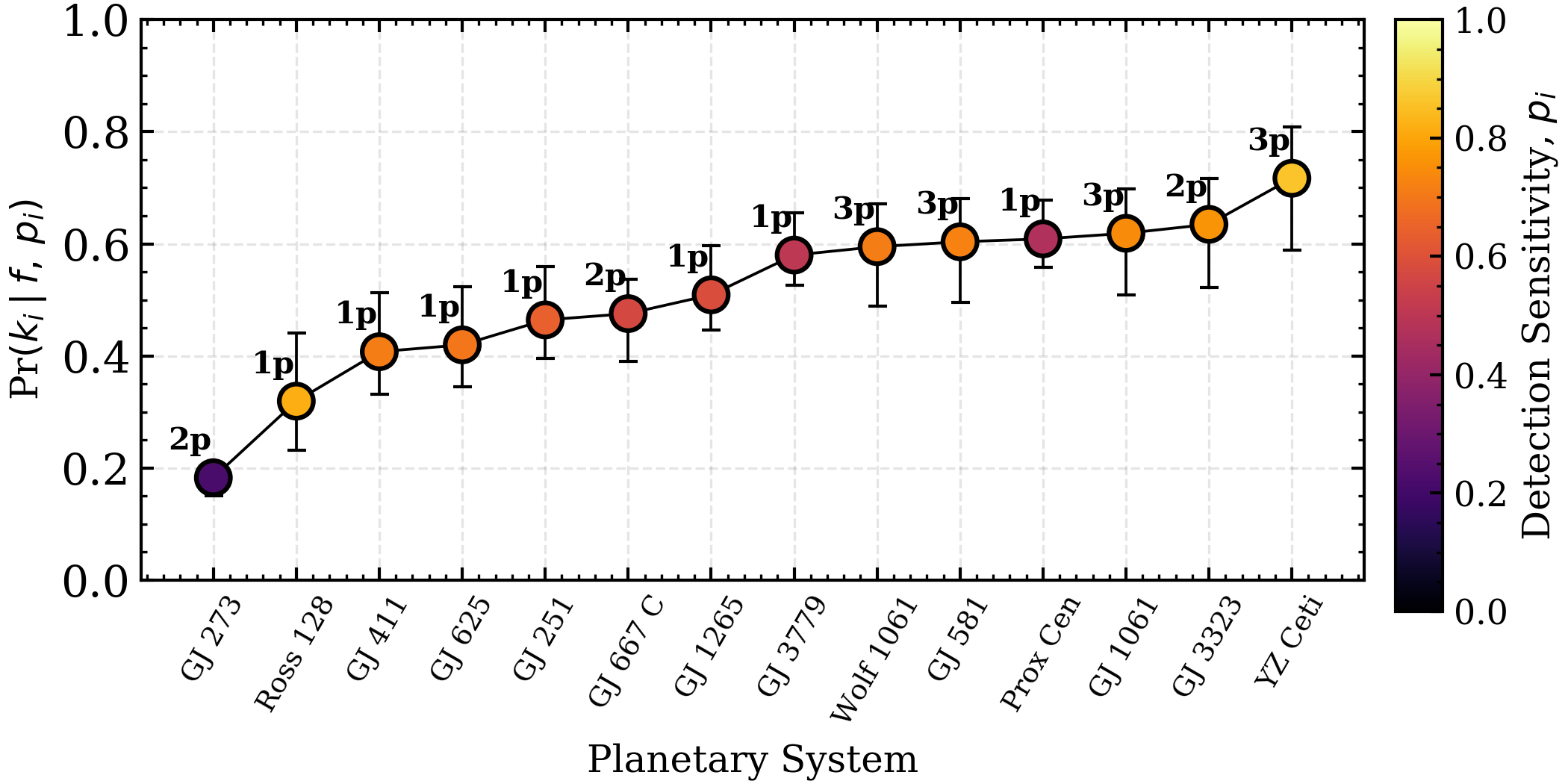}
  \caption{The probability of detecting the either one (i.e. $k_i=0$) or
    multiple planets (i.e. $k_i=1$) in each
    individual mid-M dwarf system with at least one known small planet
    candidate. Each system's probability is calculated from the fraction of
    mid-M dwarf planetary systems with multi-planet systems
    $f=$ \ftransitminuskep{} and each
    system's RV detection sensitivity to planets with $m_p \in [1,10]$ \Mearth{}
    and $P \in [0.5,50]$ days; $p_i$. The uncertainty on $f$ produces the
    errorbars on each data point. The number of known planet candidates in each
    system is annotated next to its data point.}
  \label{fig:rvprob}
\end{figure*}

\subsection{Are All Small RV Planets Detected Around Mid-M Dwarfs Real?}
In general, small planets discovered in RV datasets are more difficult to
confirm than transiting planets. This is due to the typically lower S/N of RV
signals relative to transit signals and the strong impacts that stellar
activity and uneven sampling can have on the recovery of RV signals.
As a result, small RV planets have a greater probability of being
a false positive than their transiting counterparts
\citep[e.g.][]{robertson14a,robertson14b,rajpaul16,lubin21}.

Using the value of
$f$ measured with transiting planets only (i.e. \ftransitminuskep{}), we can
calculate the probability that all of the RV planets in our sample are real
planets. Among the \Nrv{} RV systems in our sample, $\sum_i\, k_i =8$
implying that single-planet RV systems may be more common than expected
based on the results from the transiting systems where the false positive
probability is negligible. Given that the RV detection sensitivities for RV
systems are often superior to the transiting planetary systems, we would
expect to detect multiple planets in closer to 90\% of the RV systems, if those
multi-planet systems existed. The fact that only eight out of \Nrv{} (53\%) of
RV systems appear to be singles presents a discrepancy with the results from the
`clean' transiting planet sample. We find
that the probability of finding only eight out of \Nrv{} multi-planet mid-M
dwarf RV systems to be $<$ \fallrv{} at 99\% confidence. That is, mid-M dwarf
planetary systems with multiple planets are common and because most of the RV
systems with at least one known planet have typically high RV detection
sensitivity, it is improbable that such a small fraction of the RV systems would
appear to be singles. This result, coupled with the modest false positive rate
for small RV planet candidates, suggests that it is highly unlikely that all
of the small RV planet candidates in our sample are real planets because the
majority of RV systems should appear to have either zero, or at least two
planets.

We note that this statistical argument is unable to identify which RV planet
candidates are most likely to be false positives. We can however investigate
the probability of detecting $k_i$ planets in each individual system given each
$p_i$ and $f=$ \ftransitminuskep{;} i.e. Pr$(k_i|f,p_i)$.
Figure~\ref{fig:rvprob} depicts these probabilities for the \Nrv{} RV systems
using the set of $\{ p_i \}$ from MCII. The system most likely to produce its
observed result is YZ Ceti ($72^{+9}_{-13}$\%), which hosts three known planet
candidates and has the largest detection sensitivity among the RV systems
\citep[$p_{\rm YZ Ceti}=86$\%;][]{astudillodefru17c}. Conversely, GJ 273 has the
lowest probability of having its two small planet candidates detected
($18^{+2}_{-3}$\%). We caution that this does not imply that the GJ 273 planet
candidates are the most likely false positives among our sample of RV systems.
Instead, the low detection sensitivity around GJ 273 is likely the
result of residual activity cycles that are left corrected in MCII and
therefore inhibit the detection of planetary signals.

\section{Summary} \label{sect:summary}
In this paper we presented a joint transit plus RV analysis of the enveloped
terrestrial \name{} b using archival Spitzer photometry and ten years of RV
follow-up with the HARPS spectrograph. Given the RV sensitivity of our data, we
do not detect any additional planets around \name{.} We place the architecture
of the \name{} system in the context of mid-M dwarf systems by calculating the
frequency of mid-M dwarf planetary systems that host multiple planets. Our main
findings are summarized below.

\begin{itemize}
\item Using 165 HARPS RV measurements, we refine the mass of \name{} b to be
  \mplanet{} \Mearth{.} If the composition of \name{} b resembles an Earth-like
  interior with 33\% Fe and 67\% MgSiO$_3$ mass fractions and 
  surrounded by an extended H/He envelope, the envelope mass fraction needed
  to explain its mass and radius (\rplanet{} \Rearth{)} is
  $X_{\rm env}=$ \Xenv{\%}.
\item We constrain the planet's orbital eccentricity to be $<$ \ecc{} at 95\%
  confidence, which narrows its secondary eclipse window to 2.8 hours at 95\%
  confidence.
\item We compute the RV detection sensitivity to additional planets around GJ
  1214 and are able to rule out planets more massive than 3 \Mearth{} within
  10 days and planets more massive than 5 \Mearth{} within the HZ.
\item We compute the RV detection sensitivities of a set of \Ntransit{} mid-M
  dwarf transiting systems (including \name{}) and measure the frequency of
  mid-M dwarf planetary systems that host multiple planets with
  $m_p \in [1,10]$ \Mearth{} and
  $P \in [0.5,50]$ days to be $f=$ \ftransitminuskep{.}
\item Given $f$, there is a \fonep{\%} chance that \name{} is a bona-fide
  single-planet system. This value is consistent with 50\% because the
  RVs presented in this paper are not very constraining and a second planet
  around GJ 1214 may be present but is beyond our current RV detection
  capabilities.
\item Also, given $f$ and the typically high detection sensitivities of
  \Nrv{} RV systems around mid-M dwarfs, the probability that all of the RV
  planets within those systems are real is just $<$ \fallrv{} at 99\% confidence.
\end{itemize}

\acknowledgements
R.C. acknowledges support from the Banting Postdoctoral Fellowship program, 
administered by the Government of Canada, and from 
the National Aeronautics and Space Administration
in support of the TESS science mission. This material is based upon work
supported by the National Aeronautics and Space Administration under grant No.
80NSSC18K0476 issued through the XRP Program.

N.A.D. acknowledges the support of FONDECYT project 3180063.

Based on observations collected at the European Southern Observatory under ESO
programmes 183.C-0972, 283.C-5022, 1102.C-0339, 183.C-0437, and 198.C-0838.

Based on data obtained from the ESO Science Archive Facility under request
number 578670.

This work is based in part on archival data obtained with the Spitzer Space
Telescope, which was operated by the Jet Propulsion Laboratory, California
Institute of Technology under a contract with NASA.

\facilities{ESO 3.6m/HARPS, Spitzer/IRAC, STELLA/WiFSIP}

\software{\texttt{astropy} \citep{astropyi,astropyii},
  \texttt{batman} \citep{kreidberg15},
  \texttt{emcee} \citep{foremanmackey13},
  \texttt{celerite} \citep{foremanmackey13},
  \texttt{scipy} \citep{scipy},
  \texttt{TERRA} \citep{anglada12}.}

\restartappendixnumbering
\appendix
\section{\name{} Global Transit + RV Model Results} \label{app:results}
\input{gj1214results}

\section{Mid-M Dwarf RV Detection Sensitivity} \label{app:sens}

\begin{figure*}
  \centering
  \includegraphics[width=.8\hsize]{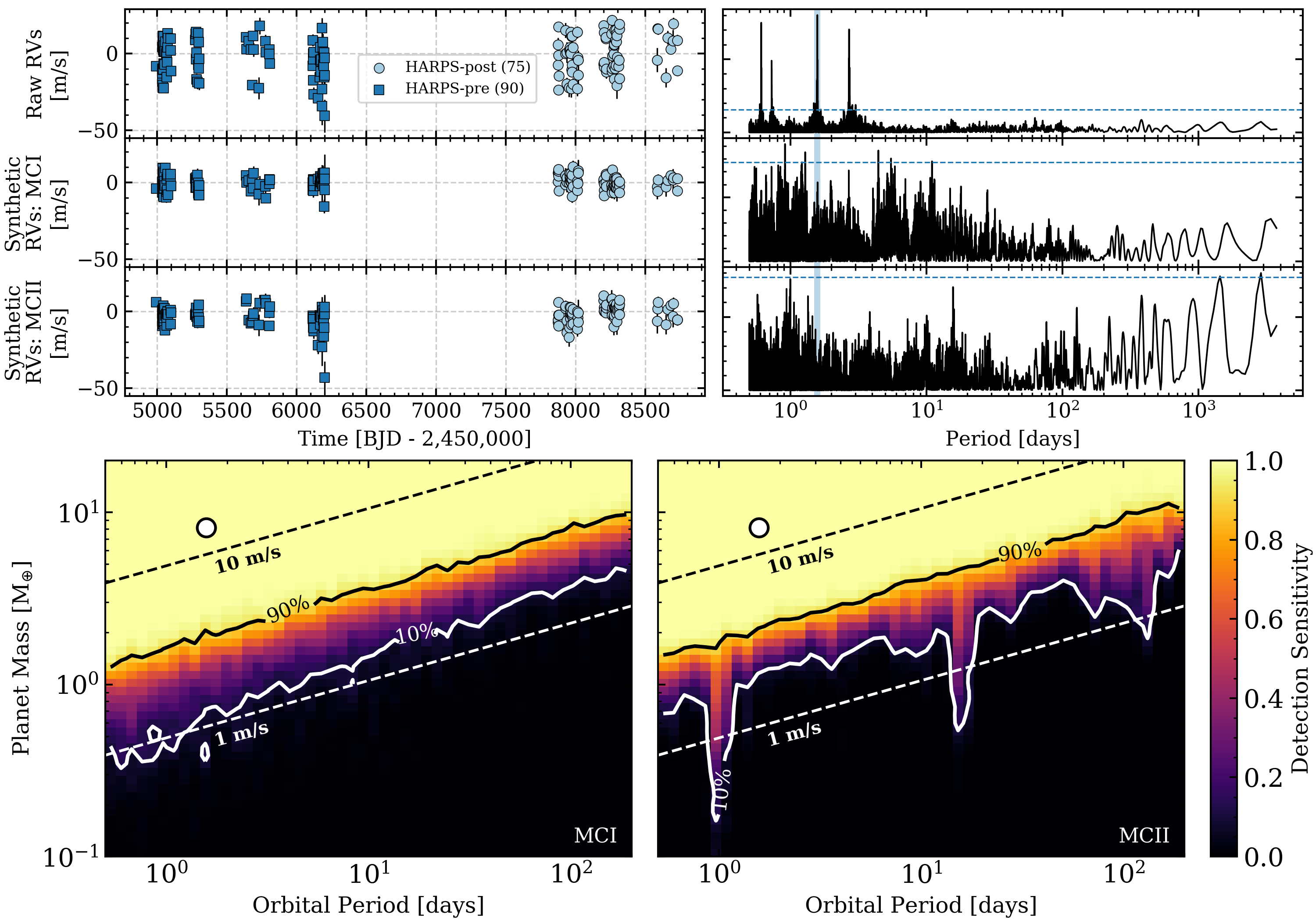}
  \caption{Raw and synthetic RVs time series, GLS periodograms, and sensitivity
    maps for GJ 1214. Top row: raw RV data from HARPS with 90 pre (squares) and
    75 (circles) post-fiber upgrade measurements. Second row: an example of a
    noisy synthetic RV dataset from the
    Monte-Carlo simulation MCI wherein the noisy RVs are constructed
    from Gaussian noise with a dispersion equal to the rms of the RV residuals.
    Third row: noisy synthetic RV data from the Monte-Carlo simulation MCII
    wherein the noisy RVs are constructed from the RV residuals after the
    removal of GJ 1214 b only (i.e. no GP component is removed). Right column:
    each RV time series' corresponding GLS periodogram.
    The vertical shaded regions highlight the orbital period of the planet
    GJ 1214 b ($P=1.58$ days).
    The horizontal dashed lines depict the approximate 1\% FAP levels.
    Lower left panel: the RV detection sensitivity computed from MCI as a
    function of planet mass and orbital period. Lower right panel: the RV
    detection sensitivity from MCII. GJ 1214 b is depicted as a white circle
    marker in both panels. The solid contours highlight the 10\% and
    90\% sensitivity levels. The dashed lines represent lines of constant
    semiamplitude with illustrative values equal to 1 and 10 \mps{.}}
  \label{fig:app1}
\end{figure*}

\begin{figure*}
  \centering
  \includegraphics[width=.8\hsize]{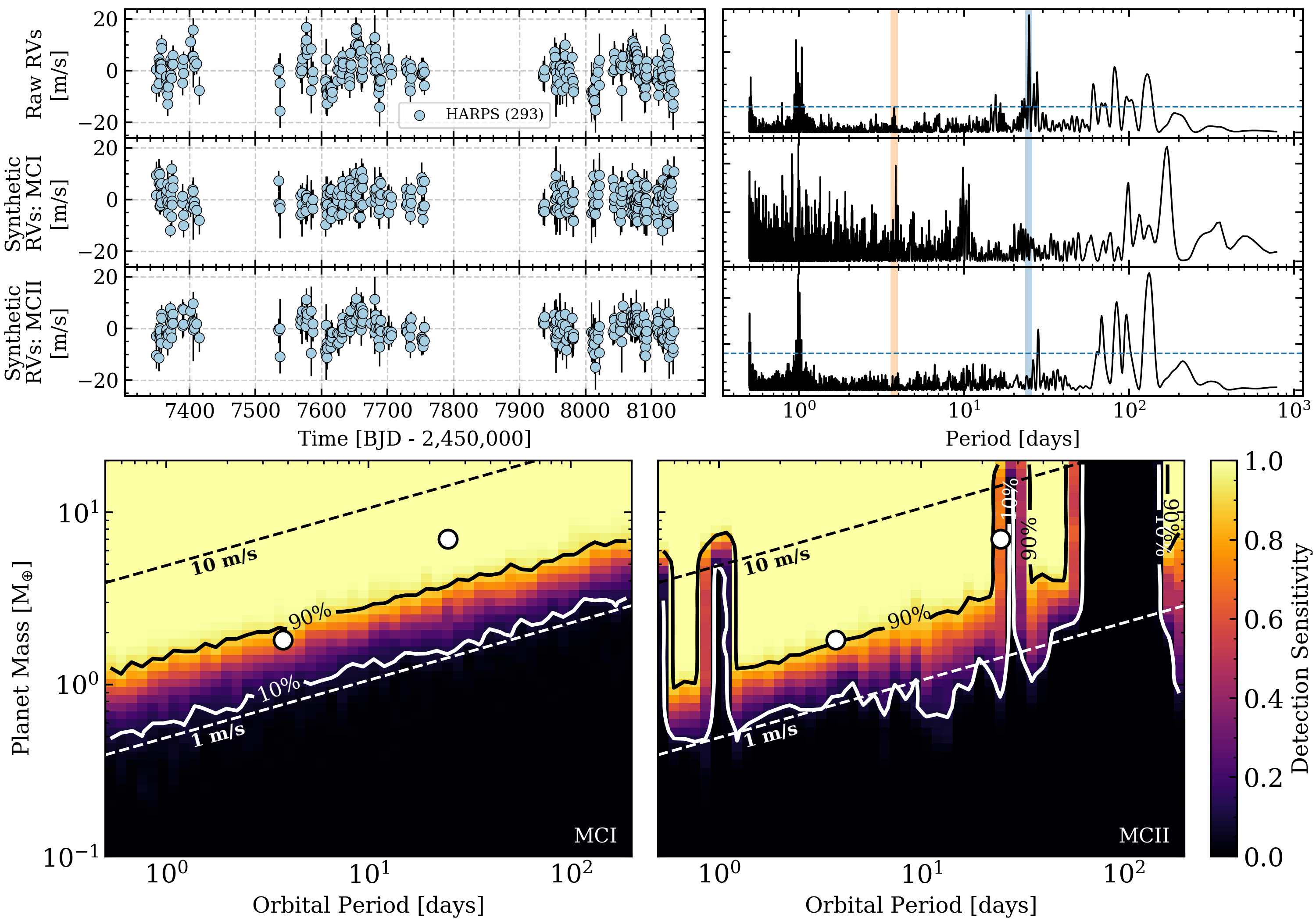}
  \caption{Similar to Figure~\ref{fig:app1} but for the LHS 1140 transiting
    system using RV data from HARPS \citep{ment19}.}
  \label{fig:app2}
\end{figure*}

\begin{figure*}
  \centering
  \includegraphics[width=.8\hsize]{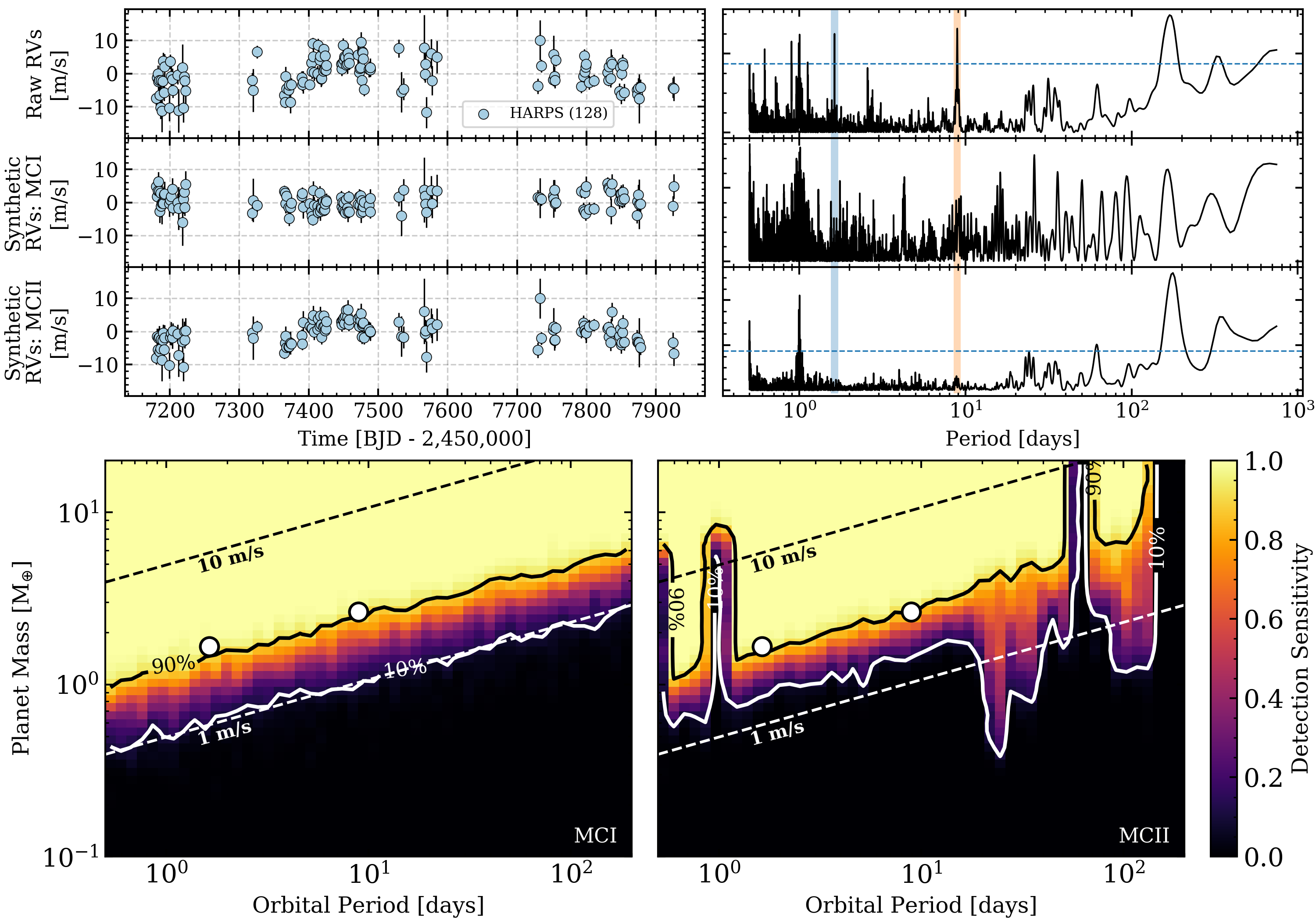}
  \caption{Similar to Figure~\ref{fig:app1} but for the GJ 1132 transiting
    system using RV data from HARPS \citep{bonfils18b}.}
  \label{fig:app3}
\end{figure*}

\begin{figure*}
  \centering
  \includegraphics[width=.8\hsize]{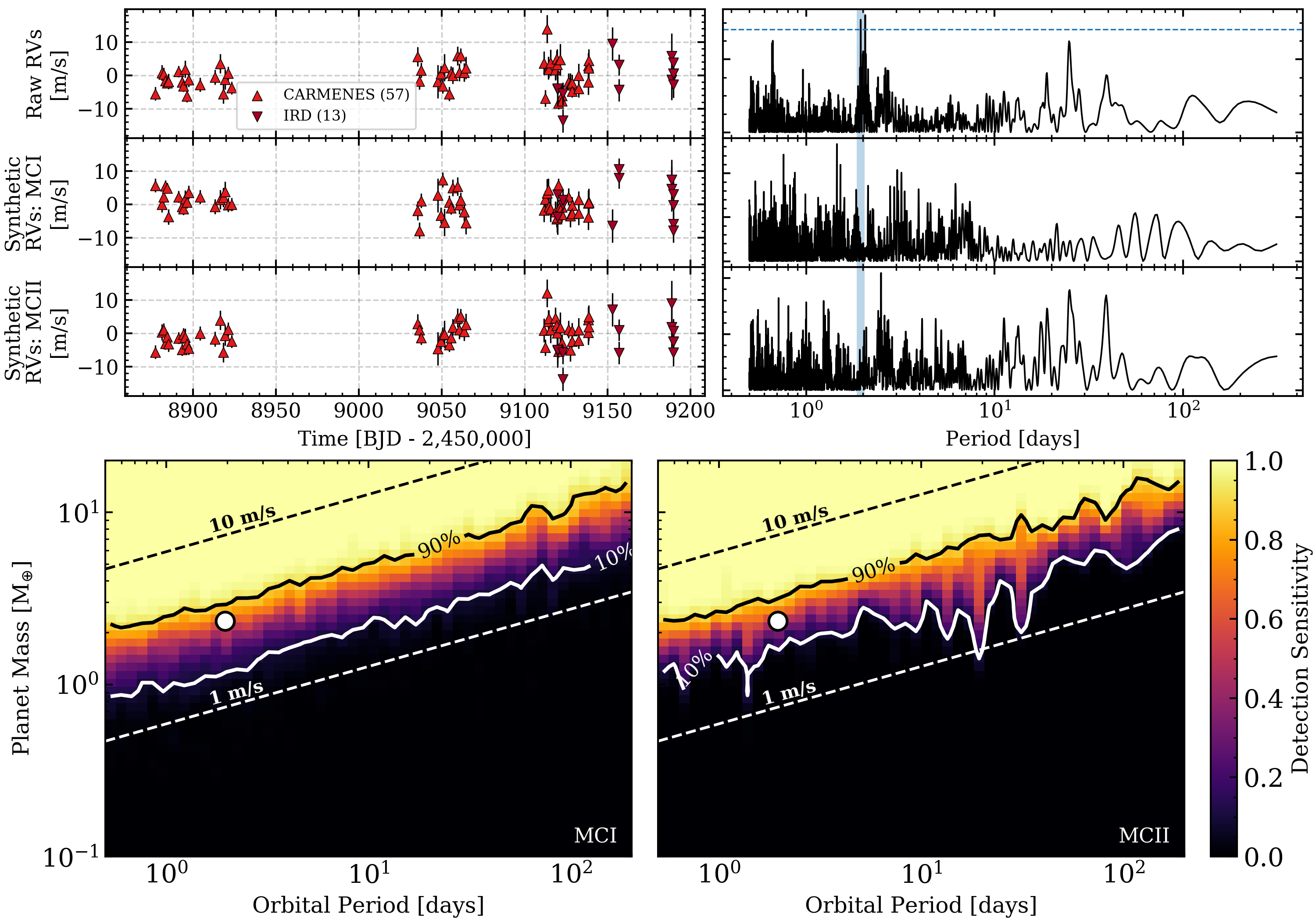}
  \caption{Similar to Figure~\ref{fig:app1} but for the LHS 1478 transiting
    system using RV data from CARMENES and IRD \citep{soto21}.}
  \label{fig:app4}
\end{figure*}

\begin{figure*}
  \centering
  \includegraphics[width=.8\hsize]{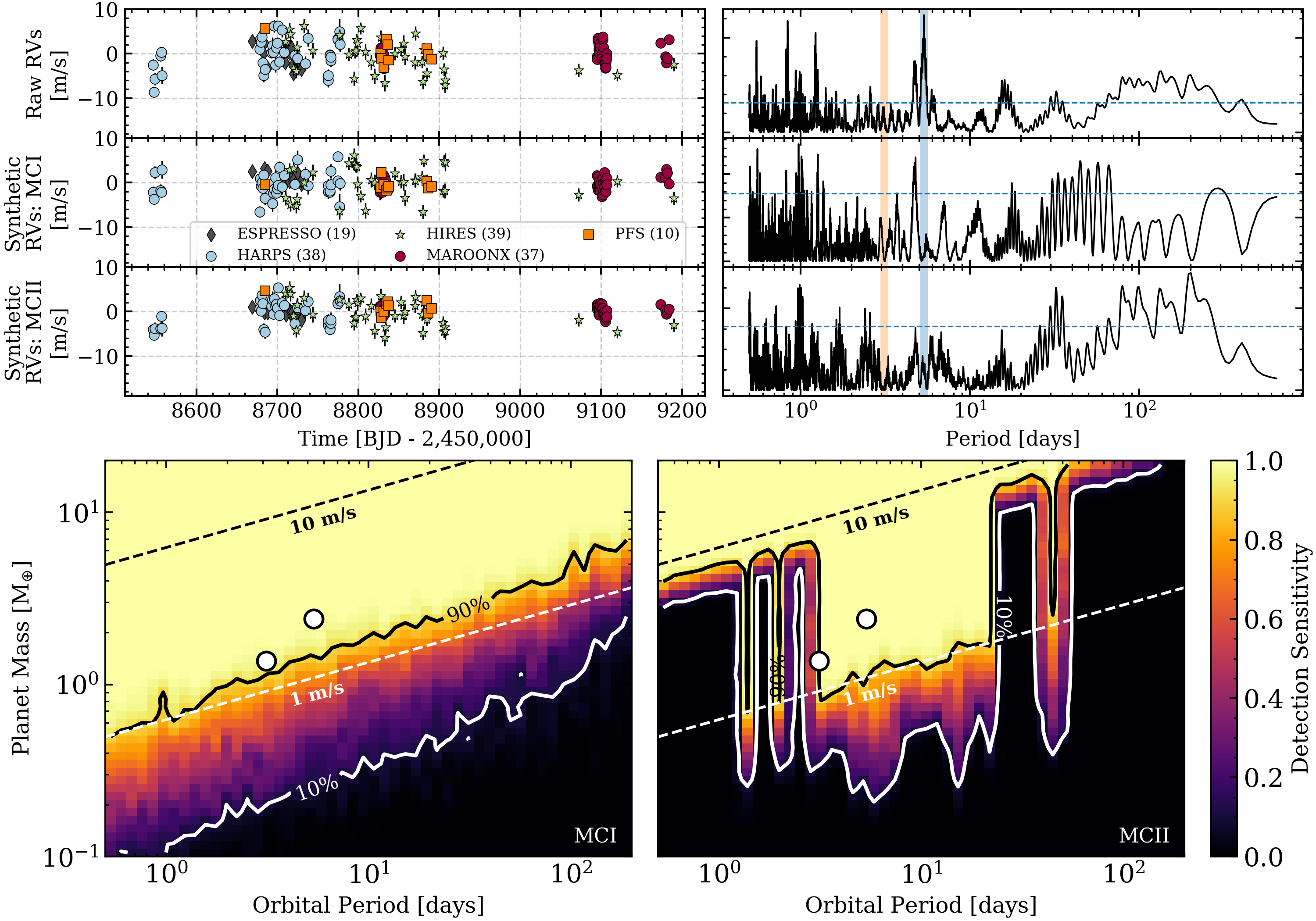}
  \caption{Similar to Figure~\ref{fig:app1} but for the LTT 1445 A transiting
    system using RV data from ESPRESSO, HARPS, HIRES, MAROON-X, and PFS
    (Winters et al. 2021 submitted).}
  \label{fig:app5}
\end{figure*}

\begin{figure*}
  \centering
  \includegraphics[width=.8\hsize]{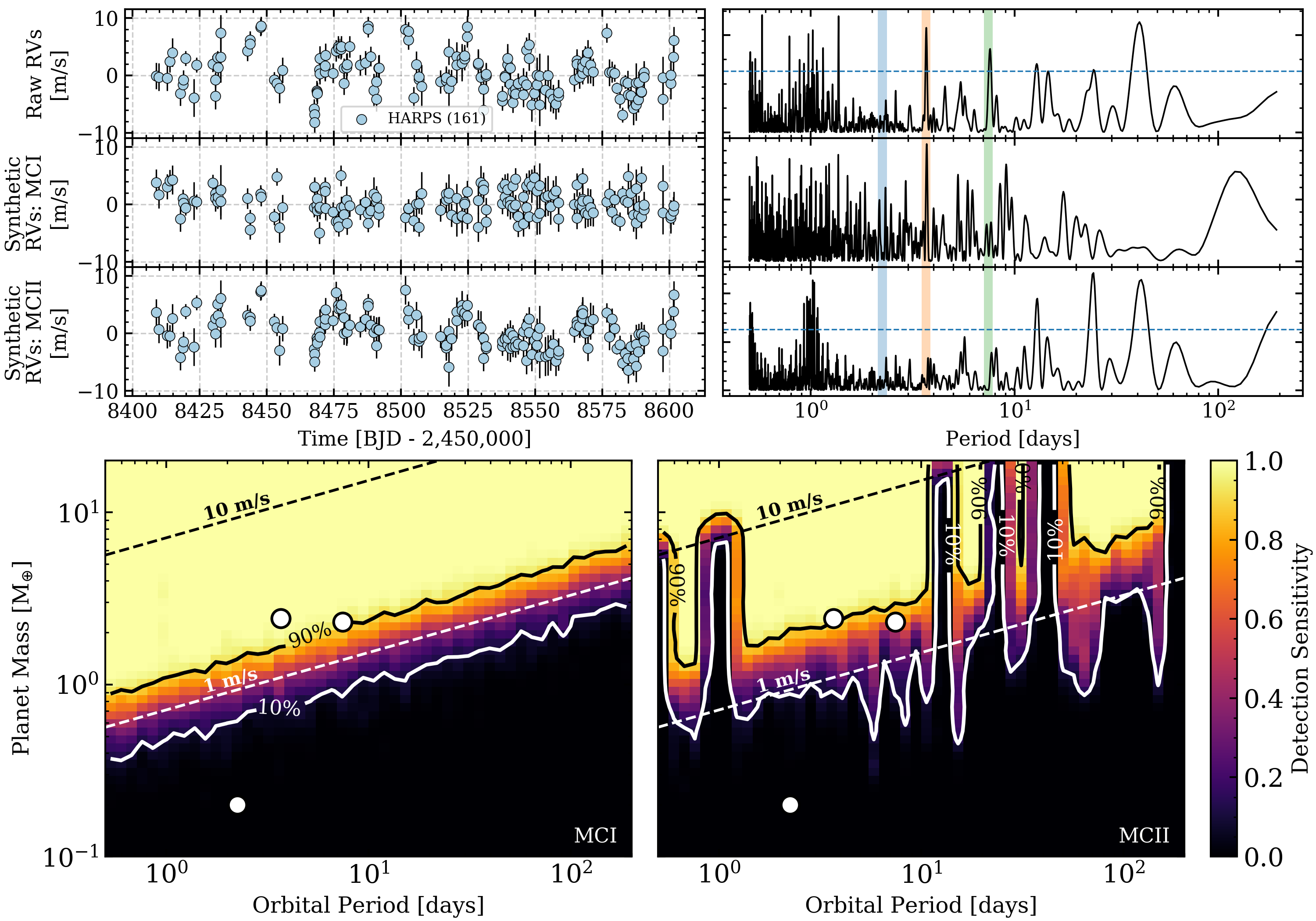}
  \caption{Similar to Figure~\ref{fig:app1} but for the L 98-59 transiting
    system using RV data from HARPS \citep{cloutier19c}.}
  \label{fig:app6}
\end{figure*}

\begin{figure*}
  \centering
  \includegraphics[width=.8\hsize]{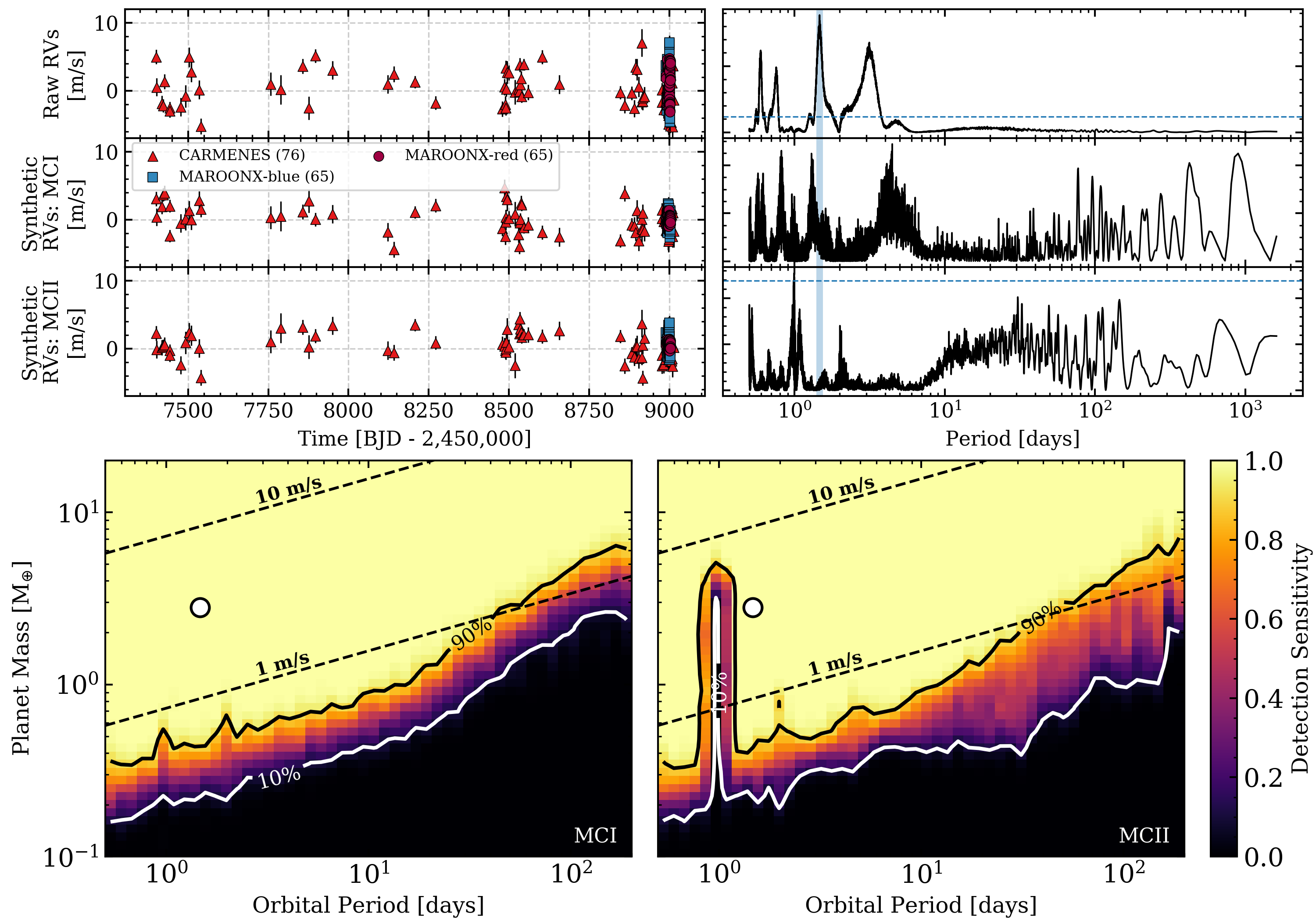}
  \caption{Similar to Figure~\ref{fig:app1} but for the GJ 486 transiting system
    using RV data from CARMENES, MAROON-X blue, and MAROON-X red
    \citep{trifonov21}.}
  \label{fig:app7}
\end{figure*}

\begin{figure*}
  \centering
  \includegraphics[width=.8\hsize]{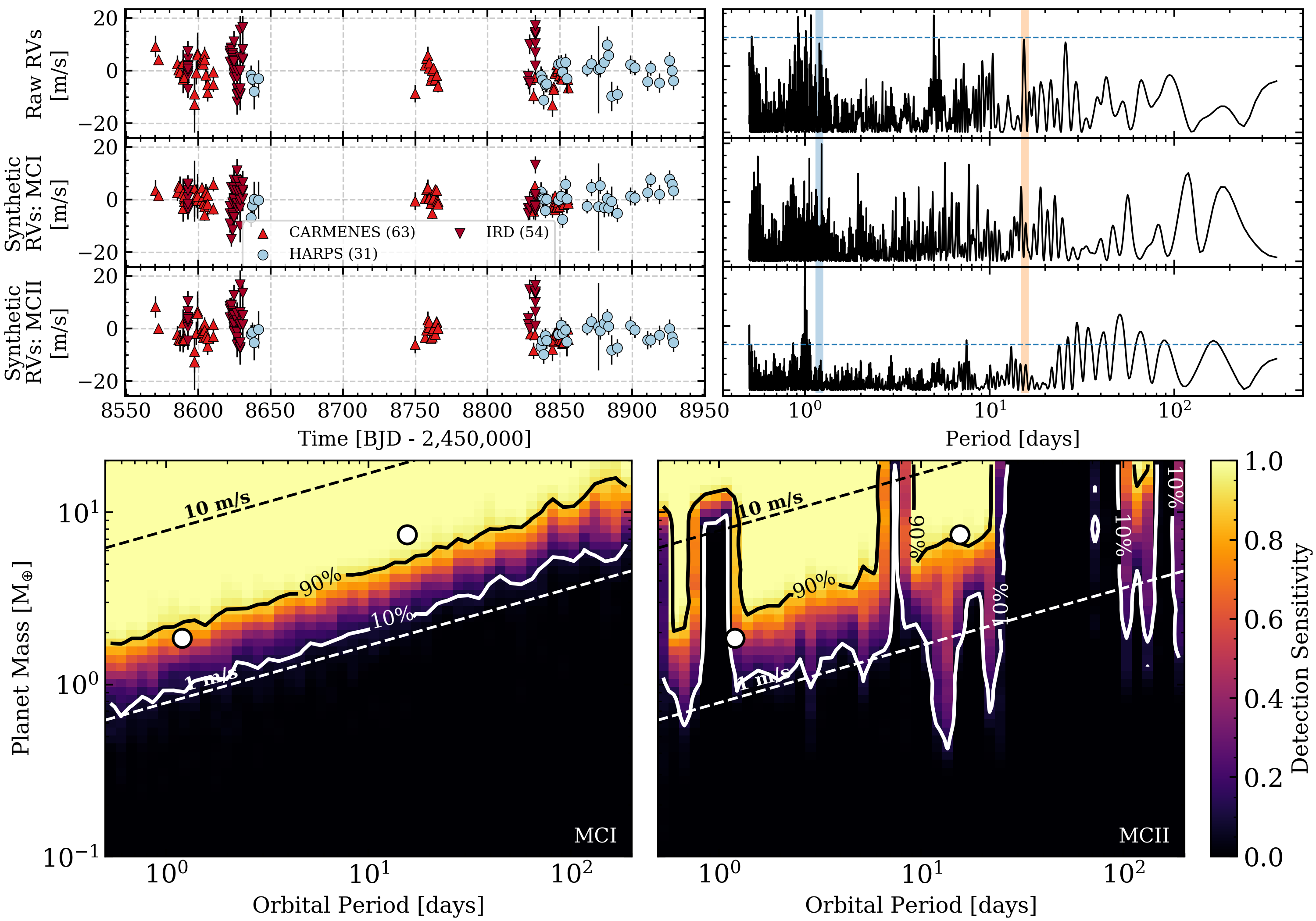}
  \caption{Similar to Figure~\ref{fig:app1} but for the GJ 3473 transiting
    system using RV data from CARMENES, HARPS, and IRD \citep{kemmer20}.}
  \label{fig:app8}
\end{figure*}

\begin{figure*}
  \centering
  \includegraphics[width=.8\hsize]{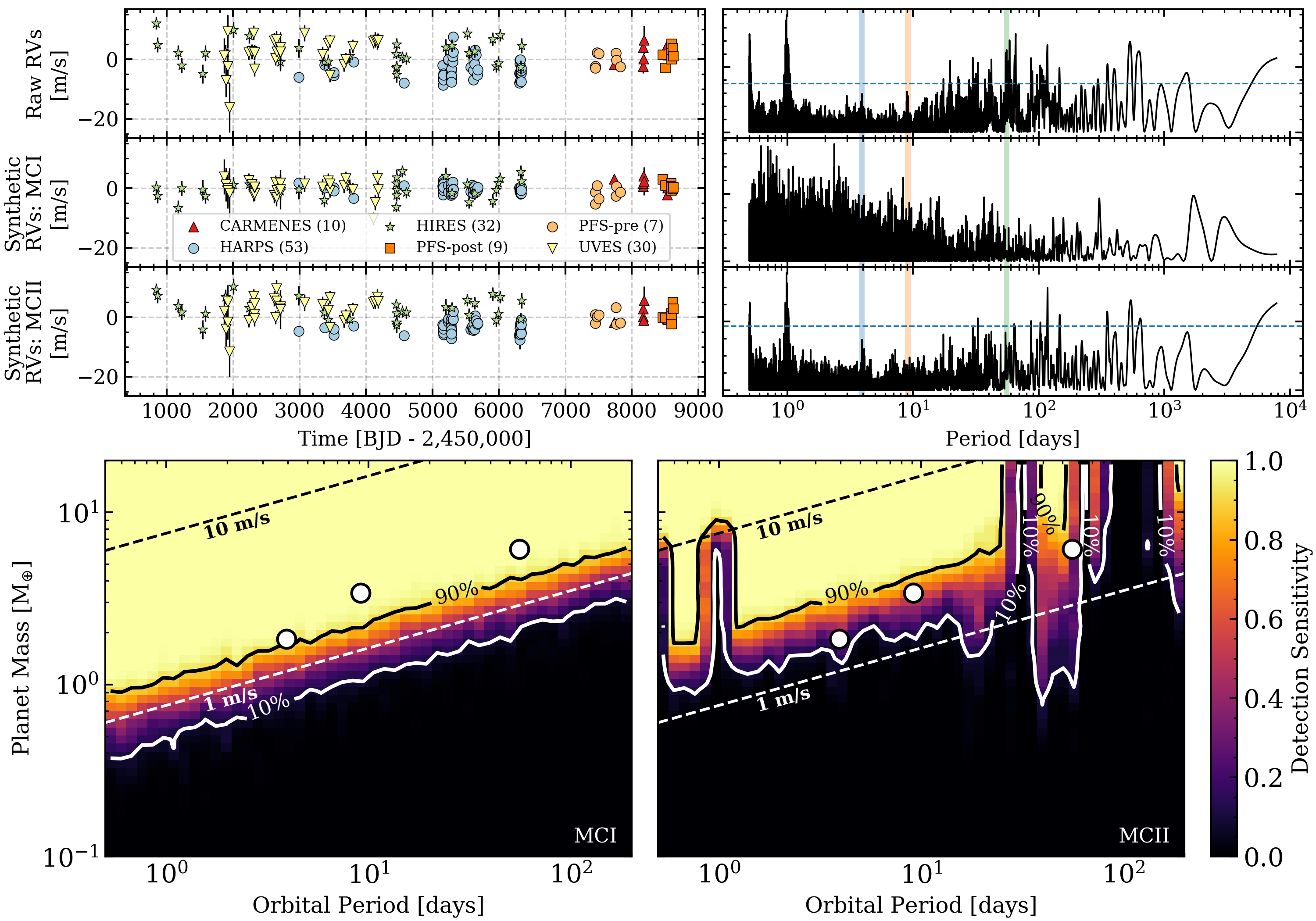}
  \caption{Similar to Figure~\ref{fig:app1} but for the GJ 357 transiting system
    using RV data from CARMENES, HARPS, HIRES, PFS (with two distinct iodine
    calibrations), and UVES \citep{luque19a}.}
  \label{fig:app9}
\end{figure*}

\begin{figure*}
  \centering
  \includegraphics[width=.8\hsize]{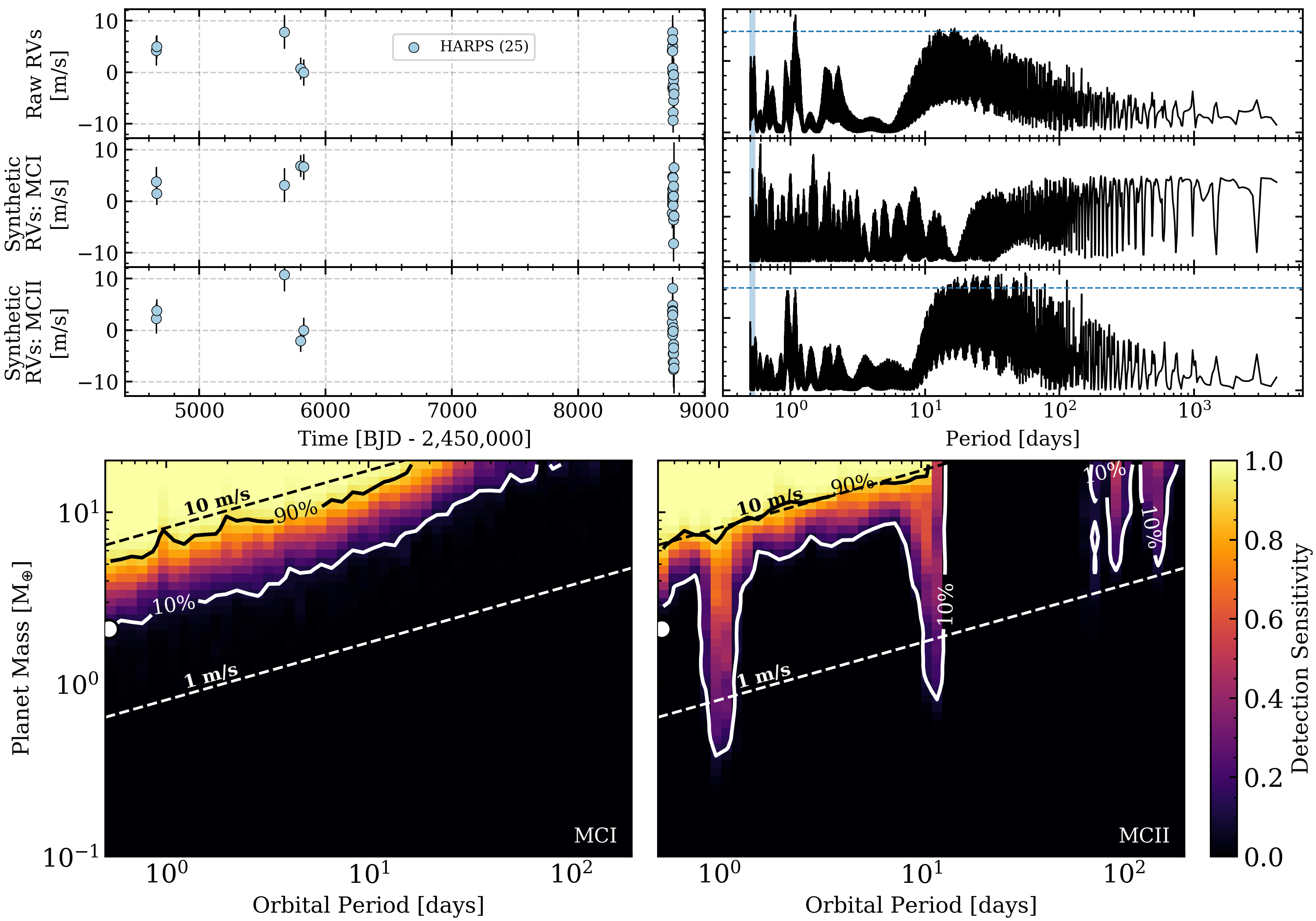}
  \caption{Similar to Figure~\ref{fig:app1} but for the GJ 1252 transiting
    system using RV data from HARPS \citep{shporer20}.}
  \label{fig:appgj1252}
\end{figure*}

\begin{figure*}
  \centering
  \includegraphics[width=.8\hsize]{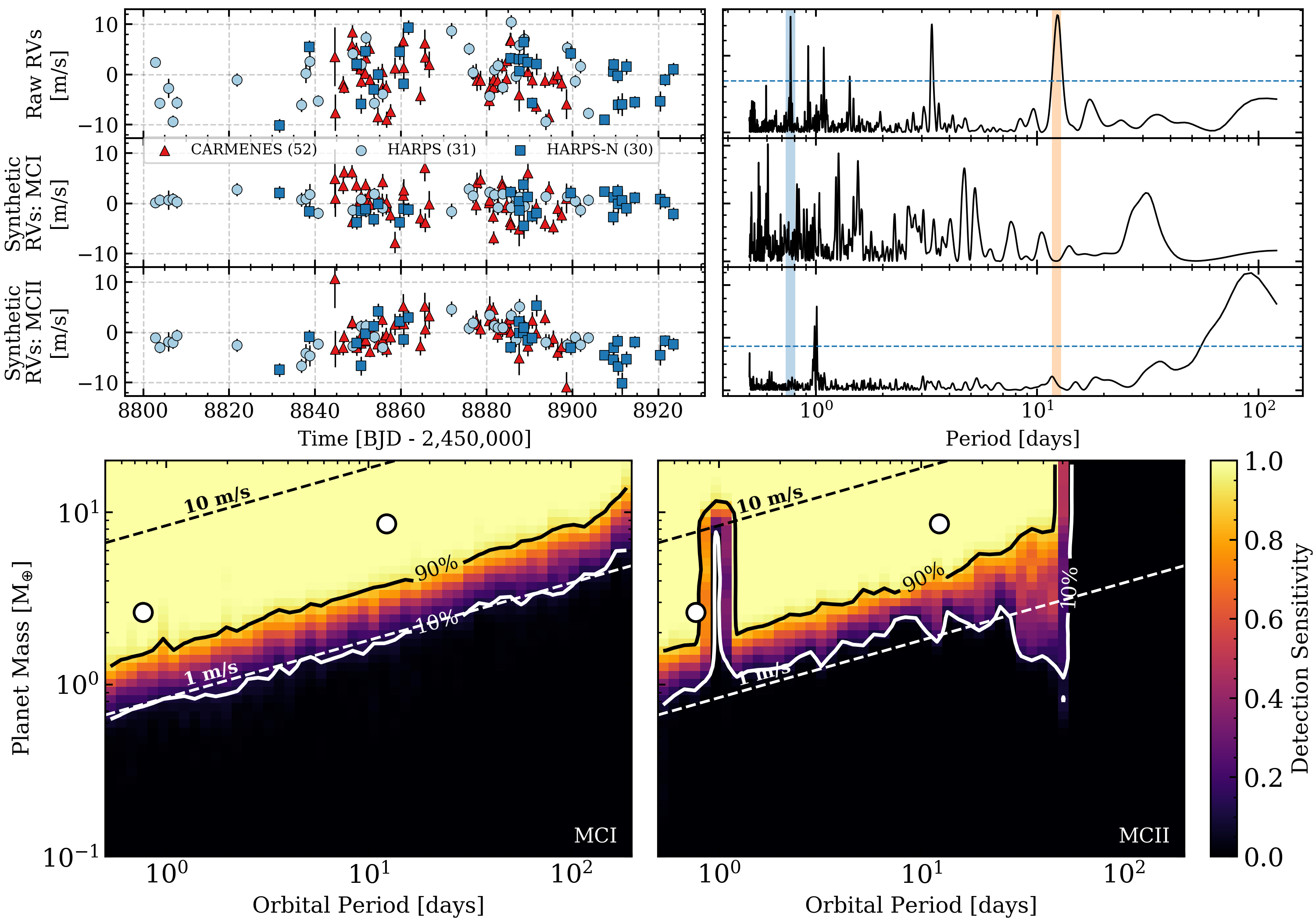}
  \caption{Similar to Figure~\ref{fig:app1} but for the LTT 3780 transiting
    system using RV data from CARMENES \citep{nowak20}, HARPS, and HARPS-N
    \citep{cloutier20b}.}
  \label{fig:app10}
\end{figure*}

\begin{figure*}
  \centering
  \includegraphics[width=.8\hsize]{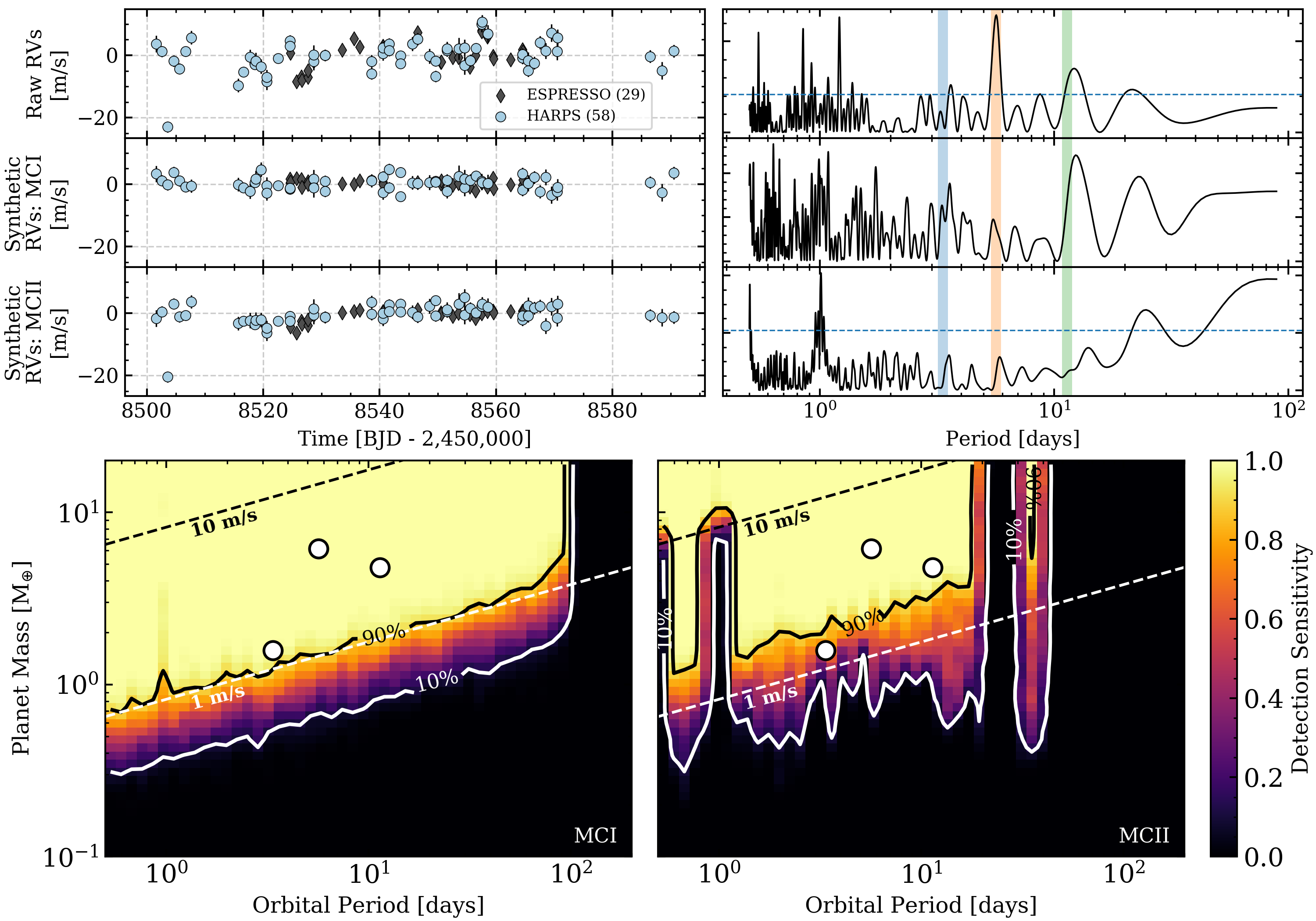}
  \caption{Similar to Figure~\ref{fig:app1} but for the TOI-270 transiting
    system using RV data from ESPRESSO and HARPS \citep{vaneylen21}.}
  \label{fig:app11}
\end{figure*}

\begin{figure*}
  \centering
  \includegraphics[width=.8\hsize]{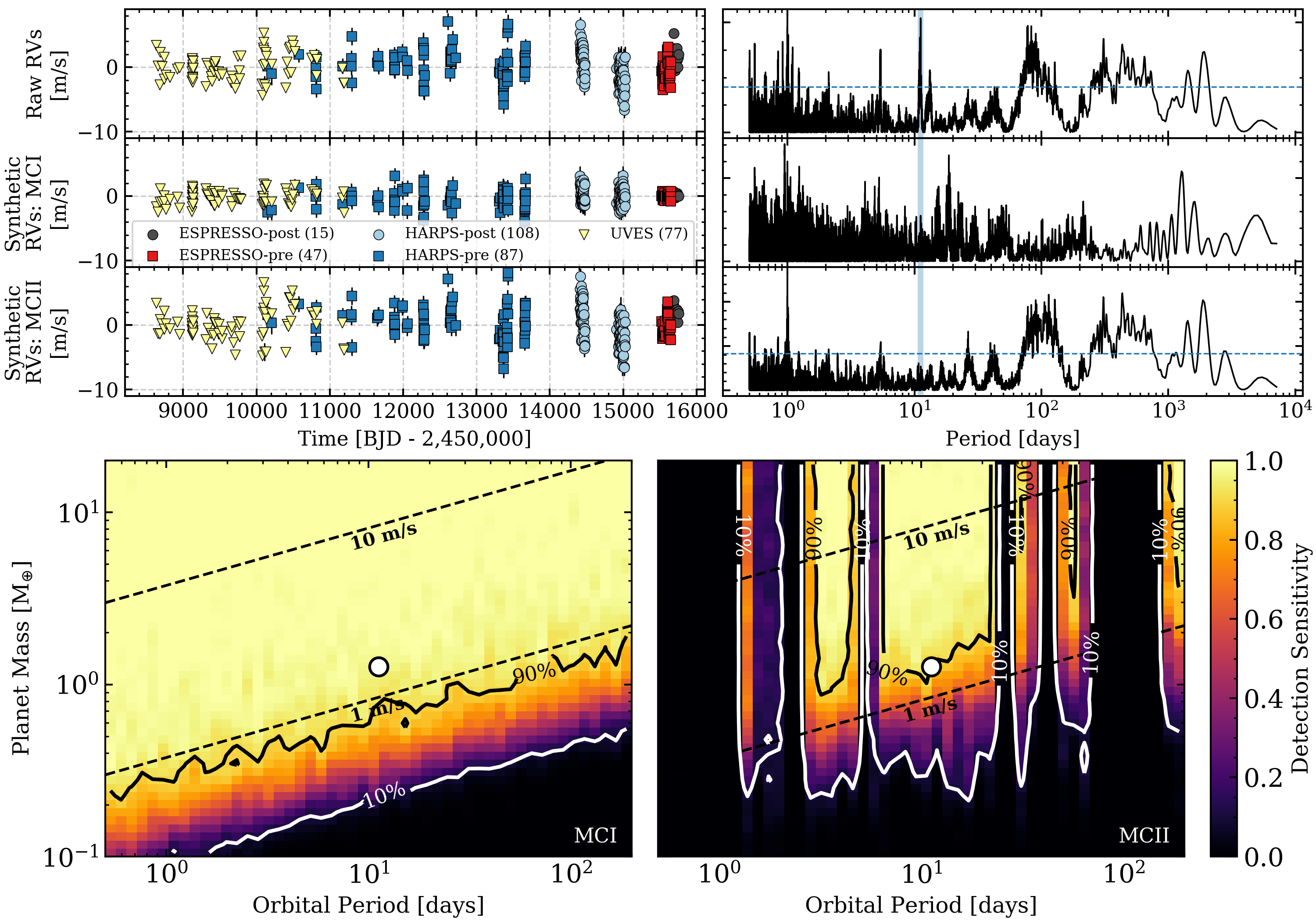}
  \caption{Similar to Figure~\ref{fig:app1} but for the Proxima Centauri RV
    system using RV data from ESPRESSO \citep{suarez20}, HARPS and UVES
    \citep{anglada16}.}
  \label{fig:app13}
\end{figure*}

\begin{figure*}
  \centering
  \includegraphics[width=.8\hsize]{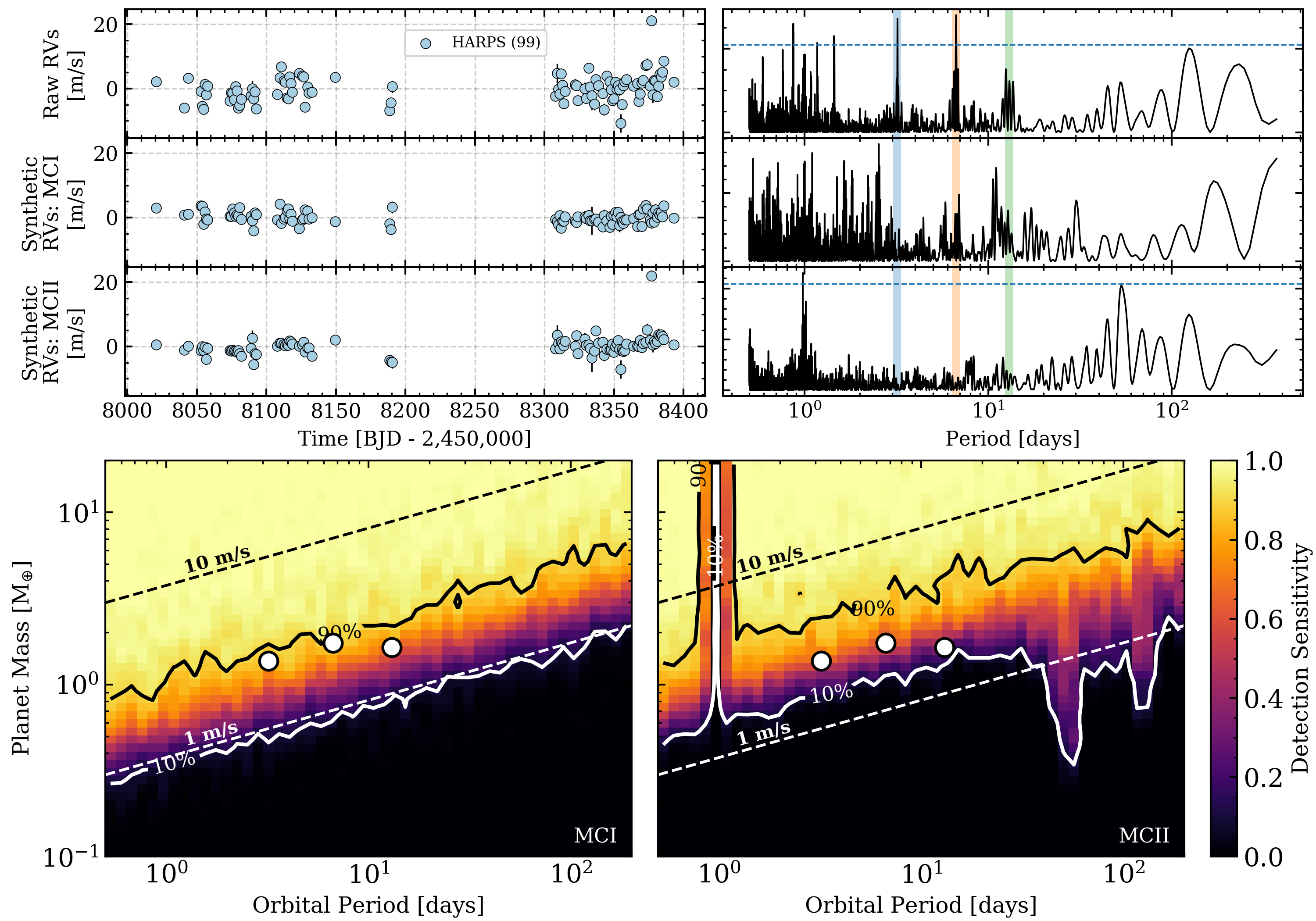}
  \caption{Similar to Figure~\ref{fig:app1} but for the GJ 1061 RV system using
    RV data from HARPS \citep{dreizler20}.}
  \label{fig:app14}
\end{figure*}

\begin{figure*}
  \centering
  \includegraphics[width=.8\hsize]{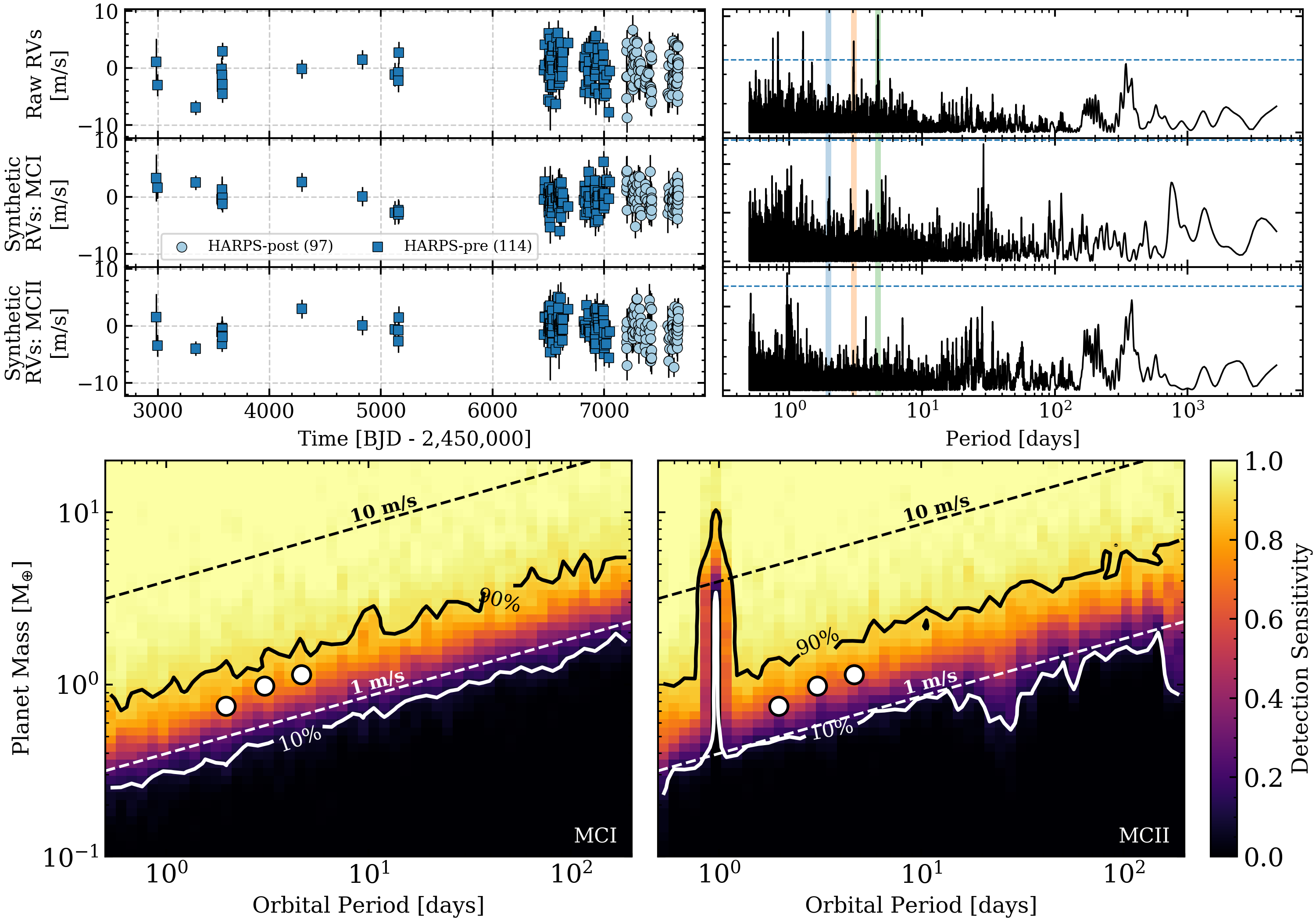}
  \caption{Similar to Figure~\ref{fig:app1} but for the YZ Ceti RV system using
    RV data from HARPS \citep{astudillodefru17c}.}
  \label{fig:app15}
\end{figure*}

\begin{figure*}
  \centering
  \includegraphics[width=.8\hsize]{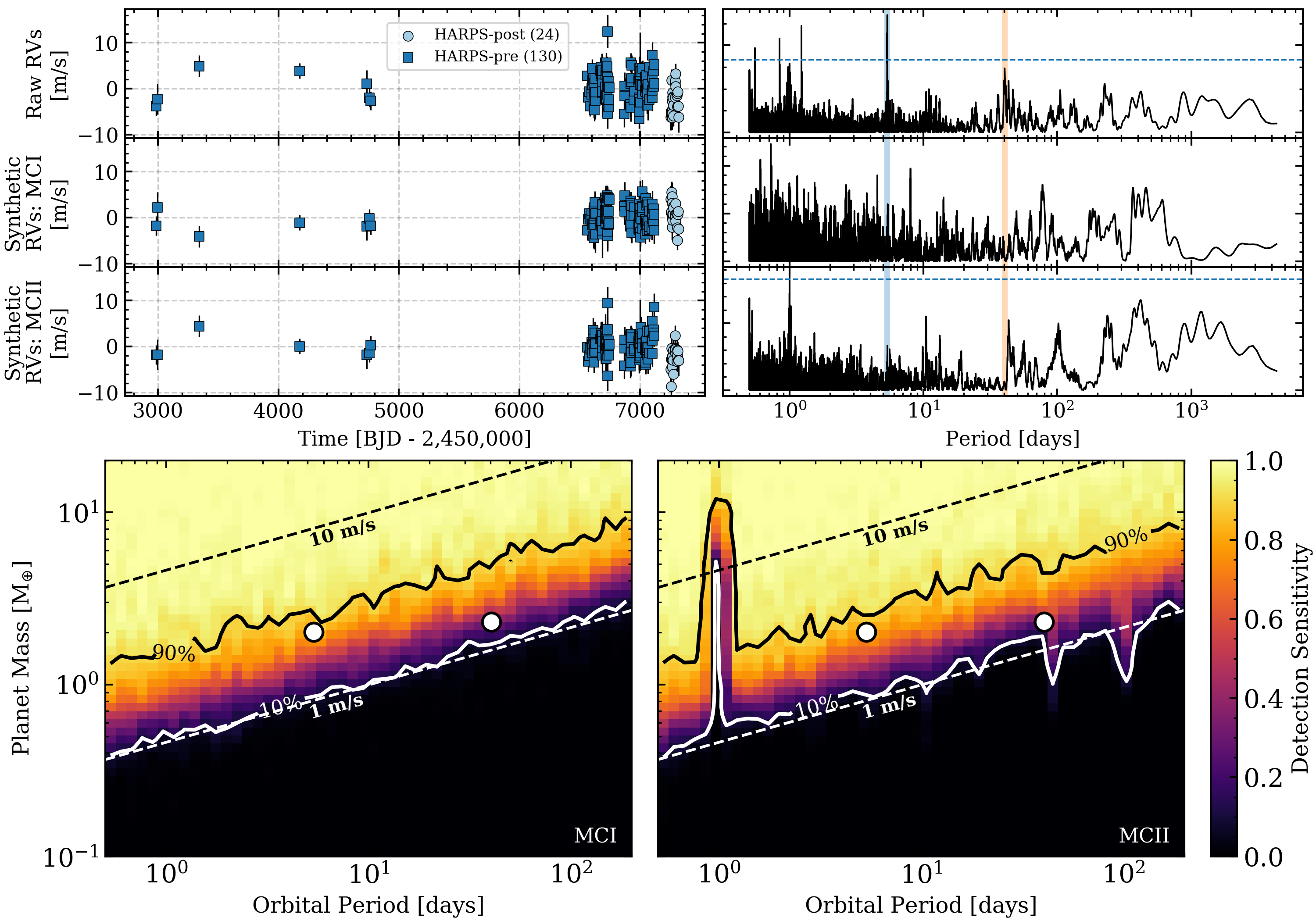}
  \caption{Similar to Figure~\ref{fig:app1} but for the GJ 3323 RV system using
    RV data from HARPS \citep{astudillodefru17a}.}
  \label{fig:app16}
\end{figure*}

\begin{figure*}
  \centering
  \includegraphics[width=.8\hsize]{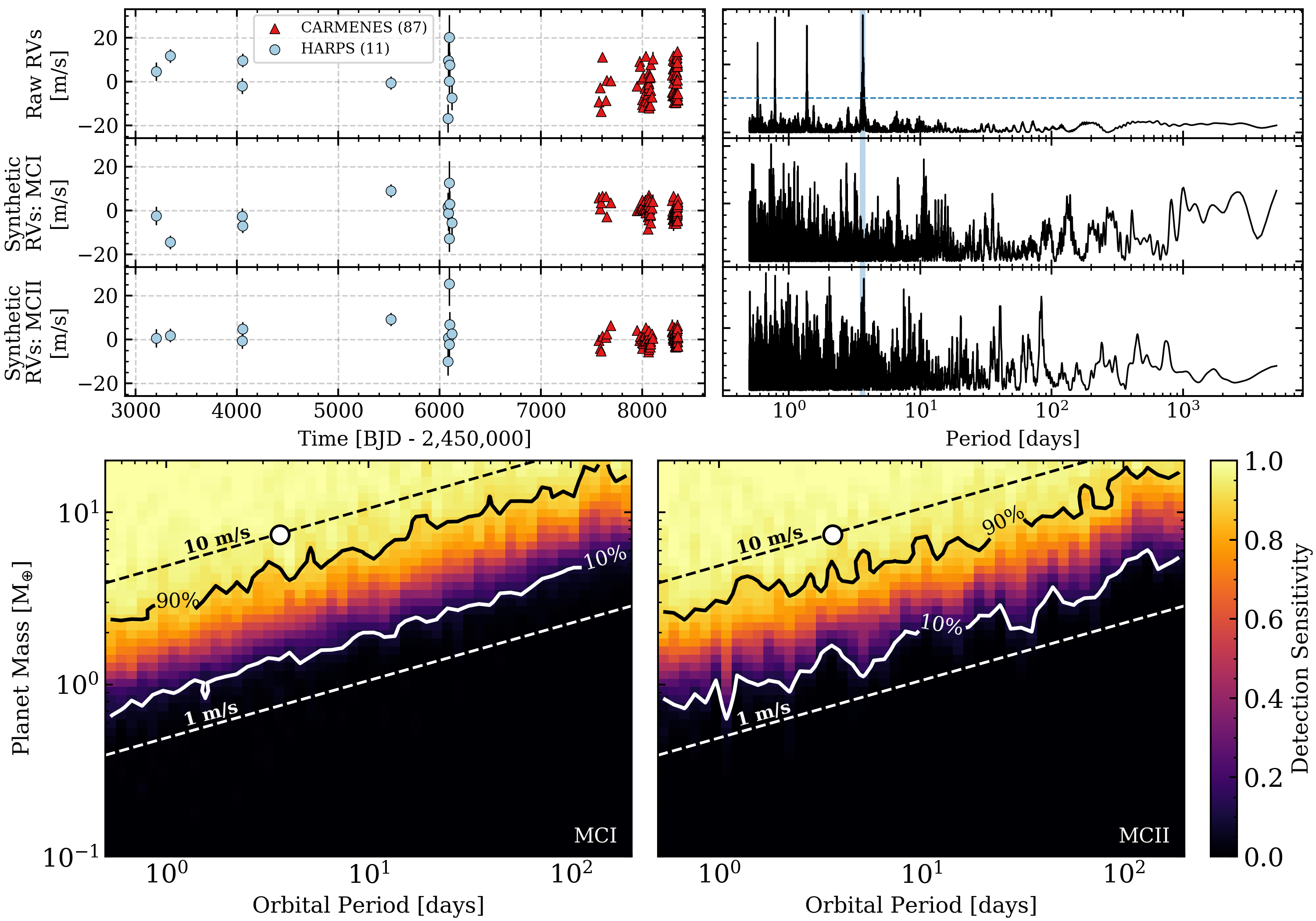}
  \caption{Similar to Figure~\ref{fig:app1} but for the GJ 1265 RV system using
    RV data from CARMENES and HARPS \citep{luque18}.}
  \label{fig:app17}
\end{figure*}

\begin{figure*}
  \centering
  \includegraphics[width=.8\hsize]{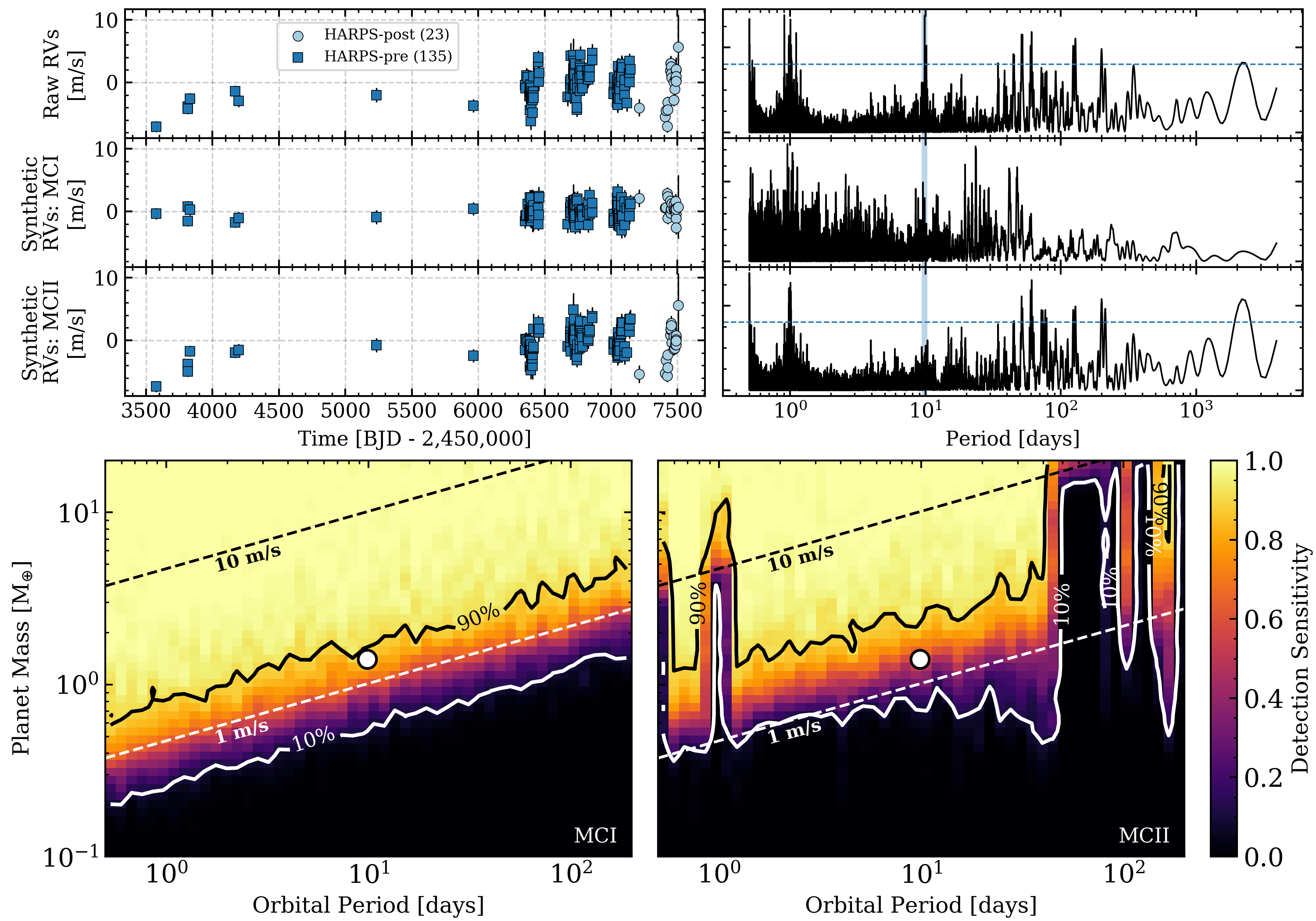}
  \caption{Similar to Figure~\ref{fig:app1} but for the Ross 128 RV system using
    RV data from HARPS \citep{bonfils18a}.}
  \label{fig:app18}
\end{figure*}

\begin{figure*}
  \centering
  \includegraphics[width=.8\hsize]{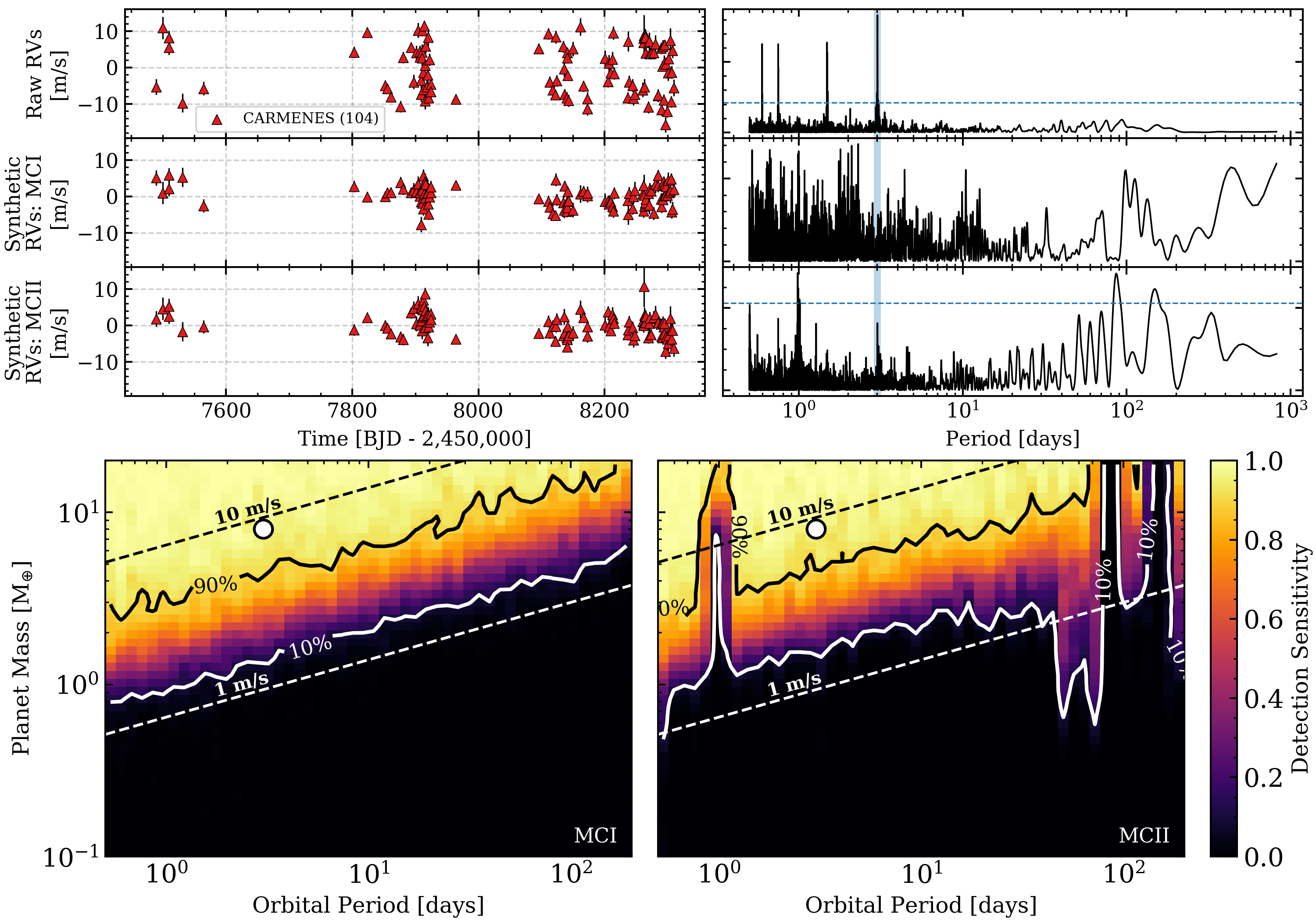}
  \caption{Similar to Figure~\ref{fig:app1} but for the GJ 3779 RV system using
    RV data from CARMENES \citep{luque18}.}
  \label{fig:app19}
\end{figure*}

\begin{figure*}
  \centering
  \includegraphics[width=.8\hsize]{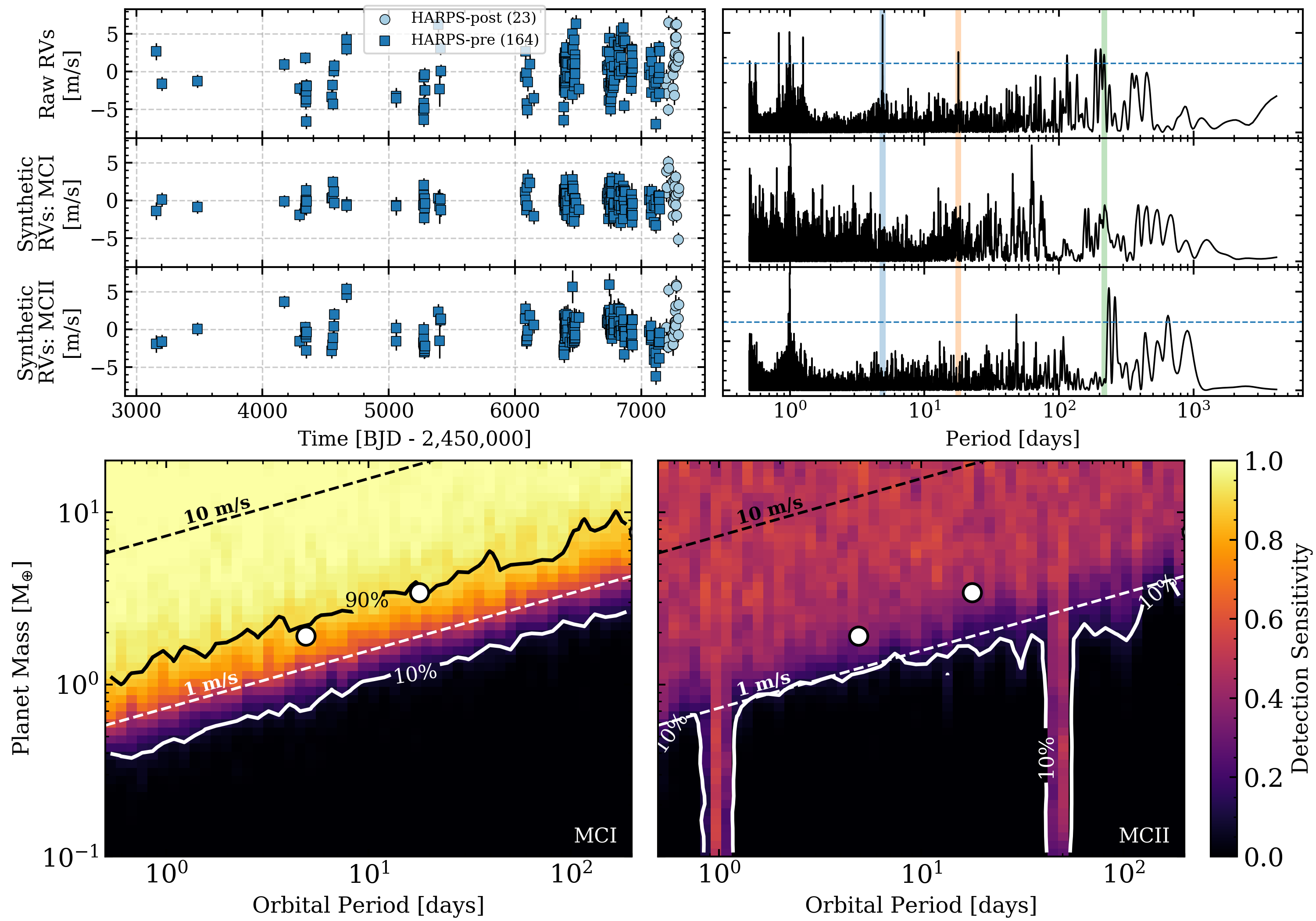}
  \caption{Similar to Figure~\ref{fig:app1} but for the Wolf 1061 RV system
    using RV data from HARPS \citep{astudillodefru17a}.}
  \label{fig:app20}
\end{figure*}

\begin{figure*}
  \centering
  \includegraphics[width=.8\hsize]{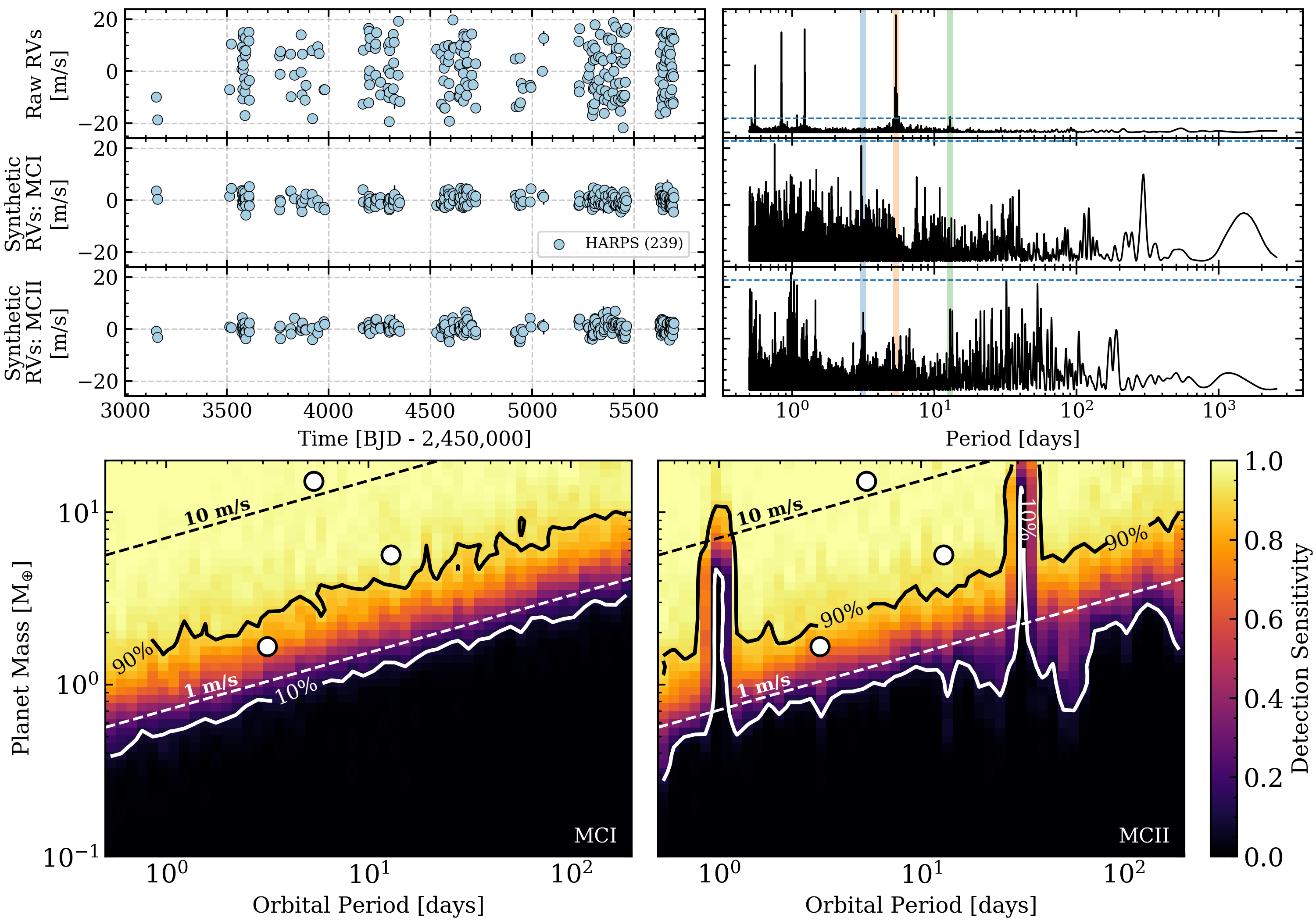}
  \caption{Similar to Figure~\ref{fig:app1} but for the GJ 581 RV system
    using RV data from HARPS \citep{trifonov18}.}
  \label{fig:appgj581}
\end{figure*}

\begin{figure*}
  \centering
  \includegraphics[width=.8\hsize]{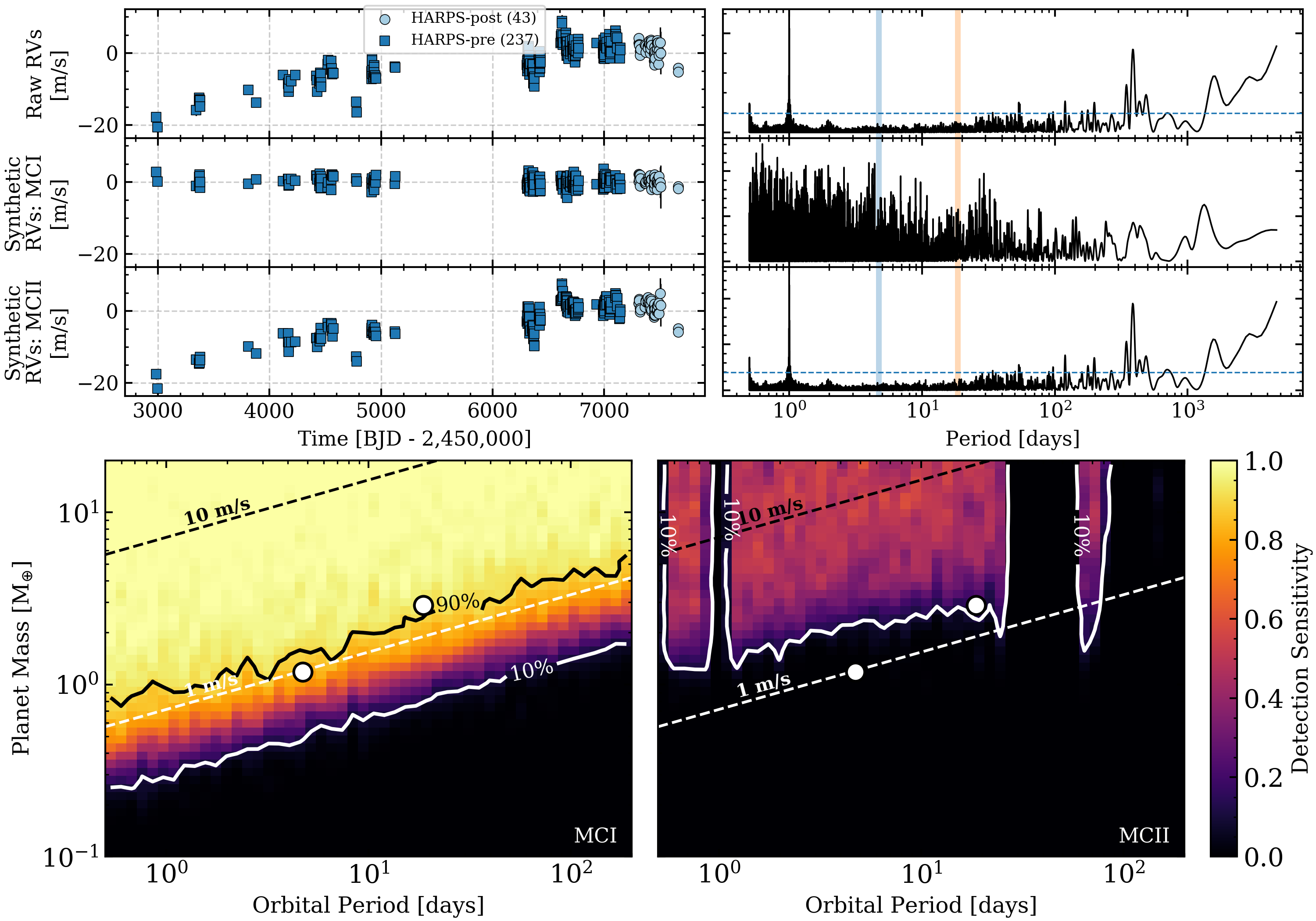}
  \caption{Similar to Figure~\ref{fig:app1} but for the GJ 273 RV system
    using RV data from HARPS \citep{astudillodefru17a}.}
  \label{fig:appgj273}
\end{figure*}

\begin{figure*}
  \centering
  \includegraphics[width=.8\hsize]{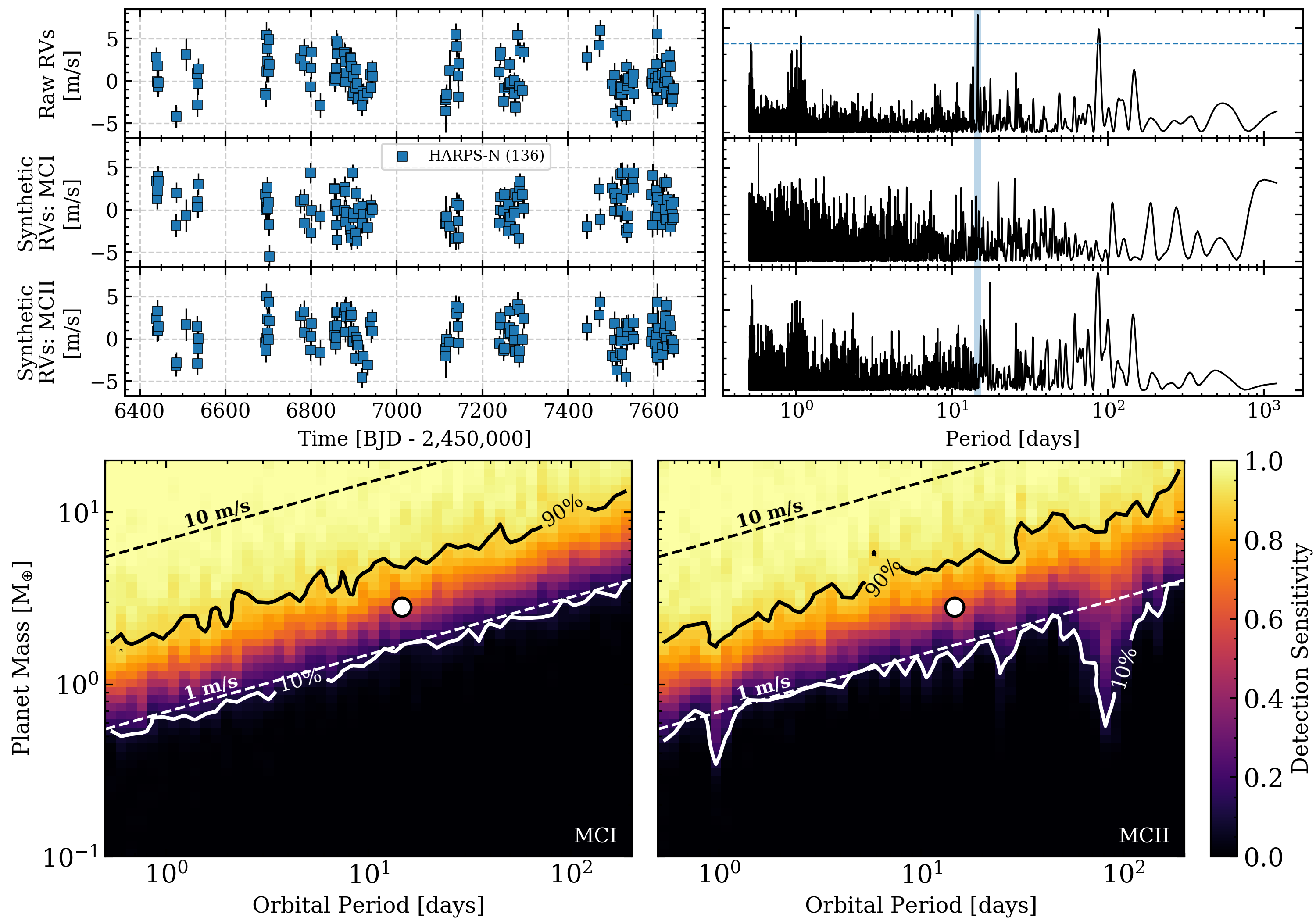}
  \caption{Similar to Figure~\ref{fig:app1} but for the GJ 625 RV system
    using RV data from HARPS \citep{suarez17}.}
  \label{fig:appgj625}
\end{figure*}

\begin{figure*}
  \centering
  \includegraphics[width=.8\hsize]{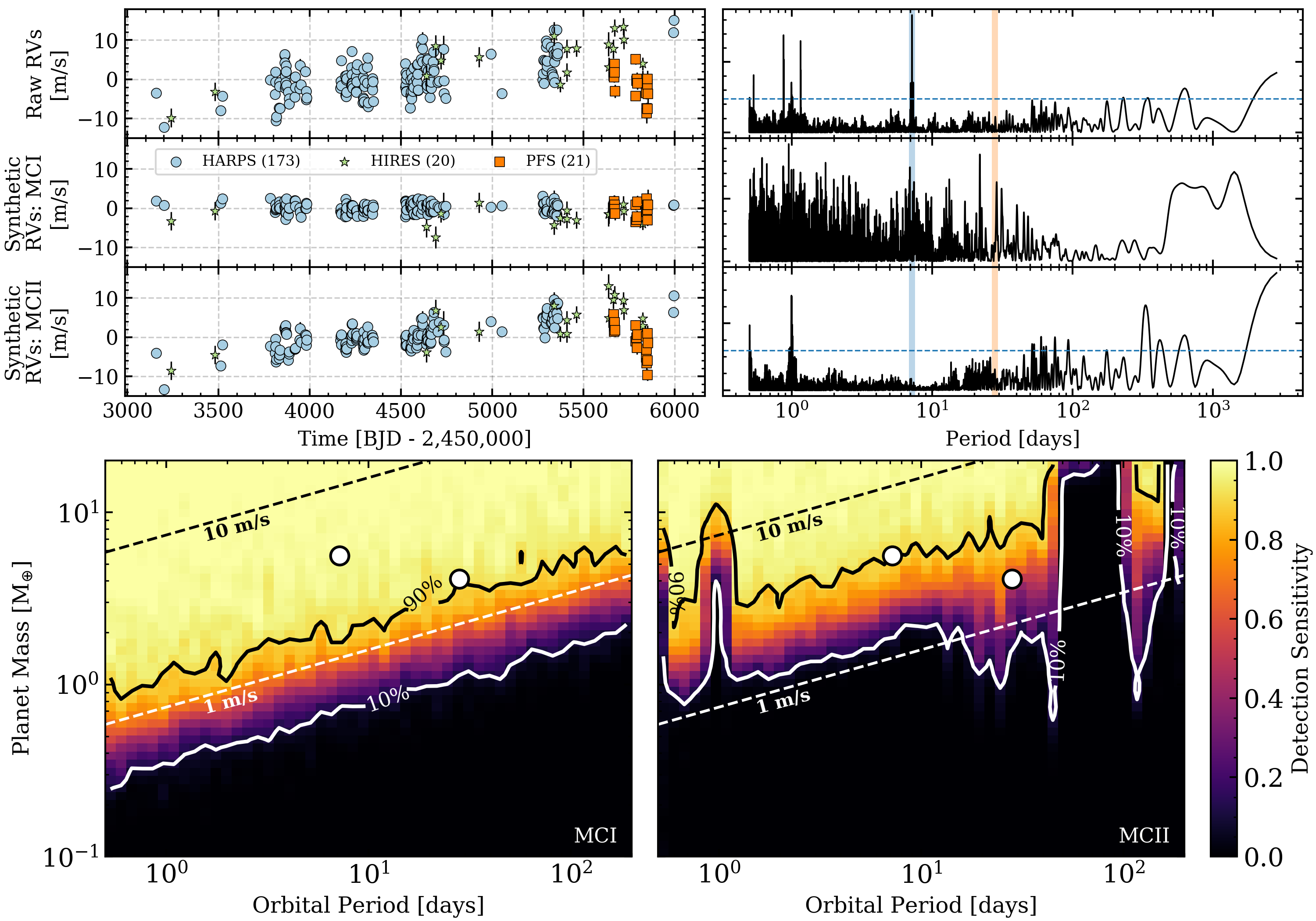}
  \caption{Similar to Figure~\ref{fig:app1} but for the GJ 667C RV system
    using RV data from HARPS, HIRES, and PFS \citep{anglada13b}.}
  \label{fig:appgj667}
\end{figure*}

\bibliographystyle{apj}
\bibliography{refs_master.bib}

\end{document}

%% file: allauthors.tex
\author[0000-0001-5383-9393]{Ryan Cloutier}
\altaffiliation{Banting Fellow}
\affiliation{Center for Astrophysics $|$ Harvard \& Smithsonian, 60 Garden
  Street, Cambridge, MA, 02138, USA}

\author[0000-0002-9003-484X]{David Charbonneau}  
\affiliation{Center for Astrophysics $|$ Harvard \& Smithsonian, 60 Garden
  Street, Cambridge, MA, 02138, USA}

\author[0000-0001-5727-4094]{Drake Deming}
\affiliation{Department of Astronomy, University of Maryland at College Park, 
  College Park, MD, 20742, USA}

\author[0000-0001-9003-8894]{Xavier Bonfils}
\affiliation{Universit\'e Grenoble Alpes, CNRS, IPAG, 38000 Grenoble, France}

\author[0000-0002-8462-515X]{Nicola Astudillo-Defru}
\affiliation{Departamento de Matem\'atica y F\'isica Aplicadas, Universidad
Cat\'olica de la Sant\`isima Concepci\'on, Alonso de Rivera 2850, Concepci\'on, Chile}

%% file: gj1214table.tex
\begin{deluxetable}{lcc}
\tabletypesize{\small}
\tablecaption{\name{} physical stellar parameters.\label{tab:star}}
\tablewidth{0pt}
\tablehead{\colhead{Parameter} & \colhead{Value} & \colhead{Ref}}
\startdata 
\multicolumn{3}{c}{\emph{GJ 1214, TIC 467929202,}} \\
\multicolumn{3}{c}{\emph{Gaia DR3 4393265392168829056}} \\
Spectral type & M4 & 1 \\
Stellar mass, $M_s$ [$M_{\odot}$] & $0.178\pm 0.010$ & 2 \\
Stellar radius, $R_s$ [$R_{\odot}$] & $0.215\pm 0.008$ & 2 \\ 
Stellar density, $\rho_s$ [g cm$^{-3}$] & $25.4^{+3.5}_{-3.0}$ & 2 \\
Stellar luminosity, $L_s$ [L$_{\odot}$] & $0.0046^{+0.0007}_{-0.0006}$ & 2 \\
Effective temperature, \teff{} [K] & $3250\pm 100$ & 3 \\
Surface gravity, \logg{} [dex] & $5.026\pm 0.040$ & 2 \\
Metallicity, [Fe/H] [dex] & $0.29\pm 0.12$ & 1 \\
\enddata
\tablecomments{\textbf{References:}
  1) \citealt{newton14}
  2) this work
  3) \citealt{anglada13a}.}
\end{deluxetable}

%% file: rvs.tex
\begin{deluxetable}{cccc}
\tablecaption{The raw HARPS RV time series of GJ 1214 in the solar system barycentric reference frame\label{tab:rv}}
\tablewidth{0pt}
\tablehead{Time & RV & $\sigma_{\text{RV}}$ & Fiber Upgrade \\
$[$BJD - 2,450,000$]$ & $[\text{m s}^{-1}]$ & $[\text{m s}^{-1}]$ & Status}
\startdata
4993.767485 & -8.239 & 2.275 & pre \\
7872.894813 & -7.473 & 3.670 & post \\
\enddata
\tablecomments{For conciseness, only a subset of rows are depicted here to illustrate the table's contents. The entirety of this table is provided in the arXiv source code.}
\end{deluxetable}

%% file: training.tex
\begin{deluxetable}{lcccc}
\tabletypesize{\small}
\tablecaption{Point estimates of the GP hyperparameters from training on the STELLA photometry\label{tab:hyperparam}}
\tablewidth{0pt}
\tablehead{Hyperparameter & Value}
\startdata
\multicolumn{2}{c}{\emph{Measured hyperparameters}} \\
Log $B$-band covariance amplitude, $\ln{a_{\rm GP,B}}$ & $-0.81^{+0.31}_{-0.29}$\\
Log $V$-band covariance amplitude, $\ln{a_{\rm GP,V}}$ & $-0.59^{+0.28}_{-0.27}$\\
Log $I$-band covariance amplitude, $\ln{a_{\rm GP,I}}$ & $-0.68^{+0.43}_{-0.49}$\\
Log exponential timescale, $\ln{\lambda_{\rm GP}}$ [days] & $5.85\pm 0.21$ \\
Log coherence, $\ln{\Gamma_{\rm GP}}$ & $-0.68+^{+0.01}_{-0.02}$ \\
Log stellar rotation period, $\ln{P_{\rm rot}}$ [days] & $4.825^{+0.039}_{-0.041}$ \\
\multicolumn{2}{c}{\emph{Derived hyperparameters}} \\
Stellar rotation period, $P_{\rm rot}$ [days] & \protval{} \\
\enddata
\end{deluxetable}

%% file: midtolatetransiting.tex
\begin{deluxetable}{lcccc}
\tabletypesize{\small}
\tablecaption{Summary of known mid-M dwarf\tablenotemark{a} transiting planetary systems with RV follow-up\label{tab:midtolate}}
\tablewidth{0pt}
\tablehead{Star name & $M_s$ & Number & Average RV & Ref. \\
& [\Msun{}] & of known & Sensitivity, & \\
&& planets & $p_i$ [\%]\tablenotemark{b} &}  
\startdata
GJ 1214 & $0.178$ & 1 & 65.8 & this work \\  
LHS 1140 & $0.179$ & $\geq 2$ & 75.7 & \citetalias{ment19},\citetalias{lillobox20} \\
GJ 1132 & $0.191$ & 2 & 73.4 & \citetalias{bonfils18b} \\
LHS 1478 & $0.235$ & 1 & 52.5 & \citetalias{soto21} \\
LTT 1445A & $0.257$ & 2 & 66.2 & Wi21 submitted \\
L 98-59 & $0.313$ & 3 & 57.0 & \citetalias{cloutier19c} \\ 
GJ 486 & $0.323$ & 1 & 94.8 & \citetalias{trifonov21} \\
GJ 3473 & $0.371$ & 2 & 43.3 & \citetalias{kemmer20} \\
GJ 357 & $0.380$ & 2 & 54.7 & \citetalias{luque19a} \\
GJ 1252 & $0.381$ & 1 & 14.1 & \citetalias{shporer20} \\  
LTT 3780 & $0.399$ & 2 & 60.0 & \citetalias{cloutier20b},\citetalias{nowak20} \\
TOI-270 & $0.400$ & 3 & 65.2 & \citetalias{vaneylen21} \\
\enddata
\tablenotetext{a}{In this study, mid-M dwarfs are defined as having a stellar mass $\in [0.1,0.4]$ \Msun{.}}
\tablenotetext{b}{Average RV sensitivity from MCII for the planets with $m_p \in [1,10]$ \Mearth{} and $P \in [0.5,50]$ days that follow log-uniform distributions versus $m_p$ and $P$.}
\end{deluxetable}

%% file: midtolaterv.tex
\begin{deluxetable}{lcccc}
\tabletypesize{\small}
\tablecaption{Summary of known mid-M dwarf\tablenotemark{a} RV planetary systems\label{tab:midtolateRV}}
\tablewidth{0pt}
\tablehead{Star name & $M_s$ & Number & Average RV & Ref. \\
& [\Msun{}] & of known & Sensitivity, & \\
&& planets & $p_i$ [\%]\tablenotemark{b} &}
\startdata
Prox Cen & $0.122$ & $\geq 1$ & 47.0 & \citetalias{anglada16},\citetalias{suarez20} \\  
GJ 1061 & $0.123$ & 3 & 74.3 & \citetalias{dreizler20} \\  
YZ Ceti & $0.133$ & 3 & 86.1 & \citetalias{astudillodefru17a} \\  
GJ 3323 & $0.166$ & 2 & 76.2 & \citetalias{astudillodefru17c} \\  
GJ 1265 & $0.168$ & 1 & 59.0 & \citetalias{luque18} \\  
Ross 128 & $0.173$ & 1 & 81.7 & \citetalias{bonfils18a} \\  
GJ 3779 & $0.264$ & 1 & 50.4 & \citetalias{luque18} \\  
Wolf 1061 & $0.302$ & 3 & 71.5 & \citetalias{astudillodefru17c} \\  
GJ 581 & $0.311$ & 3 & 72.5 & \citetalias{trifonov18} \\  
GJ 273 & $0.314$ & 2 & 22.0 & \citetalias{astudillodefru17c} \\  
GJ 625 & $0.320$ & 1 & 69.7 & \citetalias{suarez17} \\  
GJ 667C & $0.330$ & $\geq 2$ & 57.1 & \citetalias{anglada13b} \\  
GJ 876 & $0.350$ & 4 & 72.0 & \citetalias{trifonov18} \\  
GJ 251 & $0.360$ & 1 & 64.3 & \citetalias{stock20} \\  
GJ 411 & $0.386$ & 1 & 71.2 & \citetalias{diaz19} \\ 
\enddata
\tablenotetext{a}{In this study, mid-M dwarfs are defined as having a stellar mass $\in [0.1,0.4]$ \Msun{.}}
\tablenotetext{b}{Average RV sensitivity from MCII for the planets with \msini{} $\in [1,10]$ \Mearth{} and $P \in [0.5,50]$ days that follow log-uniform distributions versus $m_p$ and $P$.}
\end{deluxetable}

%% file: gj1214results.tex
\startlongtable
\begin{deluxetable*}{lcc}
\tabletypesize{\footnotesize}
\tablecaption{Point estimates of the GJ 1214 model parameters\label{tab:results}}
\tablewidth{0pt}
\tablehead{Parameter & Prior & Value}
\startdata
\multicolumn{3}{c}{\emph{Stellar parameters}} \\
Stellar mass, $M_s$ [\Msun{]} & $\mathcal{N}(0.178,0.010)$ & $0.178\pm 0.010$ \\
Stellar radius, $R_s$ [\Rsun{]} & $\mathcal{N}(0.215,0.008)$ & $0.215\pm 0.008$ \\
\multicolumn{3}{c}{\emph{RV parameters}} \\
Log covariance amplitude, [\mps{]} & $\mathcal{U}(-5,5)$ & $1.03^{+0.33}_{-0.35}$ \\
\hspace{12pt} $\ln{a_{\rm RV}}$ && \\
$a_{\rm RV}$, [\mps{]} & $-$ & $2.79^{+1.08}_{-0.82}$ \\
Log exponential timescale, $\ln{\lambda_{\rm GP}}$ [days] & Pr($\ln{\lambda_{\rm GP}} \vert \mathbf{f_{\rm STELLA}}$) & $5.86^{+0.23}_{-0.30}$  \\
Log coherence, $\ln{\Gamma_{\rm GP}}$ & Pr($\ln{\Gamma_{\rm GP}} \vert \mathbf{f_{\rm STELLA}}$) & $-0.69\pm 0.01$ \\
Log rotation period, $\ln{P_{\rm rot}}$ [days] & Pr($\ln{P_{\rm rot}} \vert \mathbf{f_{\rm STELLA}}$) & $4.835^{+0.038}_{-0.040}$ \\
Log jitter, $\ln{s_{\rm RV,pre}}$ [\mps{]} & $\mathcal{U}(-5,5)$ & $1.26^{+0.15}_{-0.16}$ \\
$s_{\rm RV,pre}$, [\mps{]} & $-$ & $3.52^{+0.56}_{-0.50}$ \\
Log jitter, $\ln{s_{\rm RV,post}}$ [\mps{]} & $\mathcal{U}(-5,5)$ & $0.81^{+0.28}_{-0.31}$ \\
$s_{\rm RV,post}$, [\mps{]} & $-$ & $2.25^{+0.76}_{-0.80}$ \\
Velocity offset, $\gamma_{\rm RV,pre}$ [\mps{]} & $\mathcal{U}(-\inf,\inf)$ & $-3.09^{+1.65}_{-1.66}$ \\
Velocity offset, $\gamma_{\rm RV,post}$ [\mps{]} & $\mathcal{U}(-\inf,\inf)$ & $-0.61^{+1.76}_{-1.90}$ \\
\multicolumn{3}{c}{\emph{Transit parameters}} \\
Baseline flux, $f_{0,3.6}$ & $\mathcal{U}(-\inf,\inf)$ & $2.9^{+4.6}_{-4.3} \times 10^{-5}$ \\
Baseline flux, $f_{0,4.5}$ & $\mathcal{U}(-\inf,\inf)$ & $2.1^{+2.2}_{-2.1} \times 10^{-5}$ \\
Limb darkening coefficient, $u_{1,3.6}$ & $\mathcal{N}(-0.0210,0.0052)$ & $-0.0210\pm 0.0052$  \\
Limb darkening coefficient, $u_{2,3.6}$ & $\mathcal{N}(0.1852, 0.0050)$ & $0.1852\pm 0.0050$ \\
Limb darkening coefficient, $u_{1,4.5}$ & $\mathcal{N}(-0.0046, 0.0050)$ & $-0.0046\pm 0.0050$ \\
Limb darkening coefficient, $u_{2,4.5}$ & $\mathcal{N}(0.1976, 0.0030)$ & $0.1976\pm 0.0030$  \\
\multicolumn{3}{c}{\emph{GJ 1214 b parameters (measured)}} \\
Orbital period, $P$ [days] & $\mathcal{U}(1,2)$ & \porb{} \\
Time of mid-transit, & $\mathcal{U}(5700,5702)$ & $5701.413328^{+0.000066}_{-0.000059}$ \\
\hspace{12pt} $T_0$ [BJD - 2,450,000] && \\
Planet-to-star radius ratio, $(r_p/R_s)_{3.6}$ & $\mathcal{U}(0.07,0.15)$ & $0.11608\pm 0.00034$ \\
Planet-to-star radius ratio, $(r_p/R_s)_{4.5}$ & $\mathcal{U}(0.07,0.15)$ & $0.11677\pm 0.00017$ \\
Impact parameter, $b$ & $\mathcal{U}(0,1)$ & $0.325\pm 0.025$ \\
Log RV semiamplitude, $\ln{K}$ & $\mathcal{U}(-1,4)$ & $2.665^{+0.036}_{-0.037}$ \\
$h=\sqrt{e}\cos{\omega}$ & $\mathcal{U}(-1,1)$ & $-0.015^{+0.123}_{-0.135}$ \\
$k=\sqrt{e}\sin{\omega}$ & $\mathcal{U}(-1,1)$ & $-0.008^{+0.057}_{-0.053}$ \\
\multicolumn{3}{c}{\emph{GJ 1214 b parameters (derived)}} \\
Planet radius, $r_{p,3.6}$ [\Rearth{]} & - & \rplanett{} \\
Planet radius, $r_{p,4.5}$ [\Rearth{]} & - & \rplanetf{} \\
RV semiamplitude, $K$ [\mps{]} & - & \Krv{} \\
Planet mass, $m_p$ [\Mearth{]} & - & \mplanet{} \\
Bulk density, $\rho_p$ [g cm$^{-3}$] & - & \rhoplanet{} \\
Surface gravity, $g_p$ [m s$^{-2}$] & - & $10.65^{+0.71}_{-0.67}$ \\
Escape velocity, $v_{\rm esc}$ [km s$^{-1}$] & - & $19.31^{+0.53}_{-0.54}$ \\
Scaled semimajor axis, $a/R_s$ & - & $14.85\pm 0.16$ \\
Inclination, $i$ [deg] & - & $88.7\pm 0.1$ \\
Eccentricity, $e$ & - & $<$ \ecc{}\tablenotemark{a} \\
Semimajor axis, $a$ [au] & - & \sma{} \\
Insolation, $F$ [F$_{\oplus}$] & - & \insol{} \\
Equilibrium temperature, $T_{\rm eq}$ [K]\tablenotemark{b} & - & \teqval{} \\
Envelope mass fraction, $X_{\rm env}$ [\%]\tablenotemark{c} & - & \Xenv{} \\
\enddata
\tablenotetext{a}{95\% upper limit.}
\tablenotetext{b}{Assuming uniform heat redistribution and zero albedo.}
\tablenotetext{c}{Assuming an Earth-like solid core with a 33\% iron core mass fraction.}
\end{deluxetable*}